%% file: ms.tex
\documentclass[12pt,preprint]{aastex}
\pdfoutput=1

\usepackage[utf8]{inputenc}
\usepackage[T1]{fontenc}
\usepackage[USenglish]{babel}

\usepackage{amsmath}
\usepackage{tikz}
\usepackage{pgfplots}
\pgfplotsset{compat=1.4}
\usepackage{pdflscape}
\usepackage{natbib}
\usepackage{textcomp}
\usepackage{standalone}
\usepackage{multicol}
\usepackage{ifpdf}

\usepackage[colorlinks=false,pdfborder={0 0 0}]{hyperref}
\usepackage[all]{hypcap}

\def\etal{{et~al.\null}}

\newcommand{\changed}[1]{#1}

\begin{document}
\title{Young, Star-forming Galaxies and their local Counterparts: the Evolving
Relationship of Mass--SFR--Metallicity since $z \sim 2.1$}

\author{Henry S. Grasshorn Gebhardt\altaffilmark{1,2}, 
Gregory R. Zeimann\altaffilmark{1,2}, Robin Ciardullo\altaffilmark{1,2}, 
Caryl Gronwall\altaffilmark{1,2}, Alex Hagen\altaffilmark{1,2},
Joanna S. Bridge\altaffilmark{1,2}, Donald P. Schneider\altaffilmark{1,2},
Jonathan R. Trump\altaffilmark{1,2,3}}
\altaffiltext{1}{Department of Astronomy \& Astrophysics, The Pennsylvania State
University, 525 Davey Lab, University Park, PA 16802}
\altaffiltext{2}{Institute for Gravitation and the Cosmos, The Pennsylvania
State University, University Park, PA 16802}
\altaffiltext{3}{Hubble Fellow}
\email{hsggebhardt@psu.edu, grzeimann@gmail.com, rbc@astro.psu.edu,
caryl@astro.psu.edu, hagen@psu.edu, jsbridge@psu.edu, dps@astro.psu.edu,
jtrump@psu.edu}

\setcounter{footnote}{1}

\begin{abstract}
We explore the evolution of the Stellar Mass--Star Formation Rate--Metallicity
Relation using a set of 256 COSMOS and GOODS galaxies in the redshift range
$1.90 < z < 2.35$.  We present the galaxies' rest-frame optical emission-line
fluxes derived from IR-grism spectroscopy with the {\sl Hubble Space 
Telescope\/} and combine these data with star formation rates and stellar 
masses obtained from deep, multi-wavelength (rest-frame UV to IR) photometry. 
We then compare these measurements to those for a local sample of
galaxies carefully matched in stellar mass ($7.5 \lesssim \log (M_*/M_{\odot}) 
\lesssim 10.5$) and star formation rate ($-0.5 \lesssim \log({\rm SFR}) 
\lesssim 2.5$ in $M_{\odot}$~yr$^{-1}$).  We find that the distribution of 
$z \sim 2.1$ galaxies in stellar mass-SFR-metallicity space is clearly 
different from that derived for our sample of similarly bright
($L_{{\rm H}\beta} > 3 \times 10^{40}$~ergs~s$^{-1}$) local galaxies, 
and this offset cannot be explained by simple systematic
offsets in the derived quantities.   At stellar masses above $\sim 10^9 \, 
M_{\odot}$ and star formation rates above $\sim 10 \, 
M_{\odot}$~yr$^{-1}$, the $z \sim 2.1$ galaxies have higher oxygen abundances
than their local counterparts, while the opposite is true for lower-mass,
lower-SFR systems.
\end{abstract}


%\tableofcontents

\section{Introduction}
\label{sec:intro}
Galaxy evolution depends on a host of physical processes, such as the accretion
of cold gas, the energy input from stellar winds, supernovae, and/or AGN, the
creation of dust, and the action of mergers. Each of these effects leaves its
imprint on the observed relationship between stellar mass and galactic
metallicity\footnote{Throughout this paper we loosely use the term metallicity
to refer to oxygen abundance.} \citep[e.g.,][]{lequeux+79}.  Locally, this
mass--metallicity relation has been defined by \citet{tremonti+04}, who found a
steep decline in metal abundance for low-mass objects. Interestingly, because
of the correlation between metallicity and SFR \citep[e.g.,][]{yates+12}, these
galaxies define a narrow surface in mass--metallicity--SFR space, that
is sometimes thought of as a plane \citep[e.g.,][]{lara-lopez+10}, or as a more
complicated 3D-surface \citep[e.g.]{mannucci+10, yates+12, andrews+13}.  In
this {\it Fundamental Metallicity Relation,} or FMR, metallicity and SFR are
anti-correlated at low fixed stellar mass.  At high stellar mass, this
relationship weakens, to the extent that the aforementioned references
disagree on whether the correlation between SFR and metallicity is
positive, negative, or exists at all.

Defining the relationship between stellar mass, metallicity, and star formation
rate at higher redshifts is difficult, since the rest-frame optical emission 
lines of oxygen and nitrogen that are used to determine metallicity are
shifted into sky-dominated regions of the near-infrared (NIR\null). Yet this
$1 < z < 3$ epoch is crucial to our understanding of galactic evolution, as it 
is when star formation peaked, and much of today's stellar mass and heavy 
elements were produced \citep[e.g.,][]{dickinson+03, hopkins+06}.  By comparing
the mass--SFR--metallicity (MSZ) relation of this epoch to that of the local
universe, we can test whether the correlations seen locally are 
truly fundamental (i.e., non-evolving with redshift) and place meaningful 
constraints on models of chemical evolution \citep[e.g.,][]{lilly+13, 
dayal+13}.

There have been numerous efforts aimed at measuring the $z \sim 2$ MSZ (or a
projection thereof) using a variety of bright-line metallicity indicators
including [N~II]/H$\alpha$ \citep[e.g.,][]{erb+06a, finkelstein+11, kulas+13,
zahid+14, wuyts+14, song+14, maier+14, steidel+14, sanders+15}, [O~II]/[O~III]
\citep[e.g.,][]{nakajima+13}, [O~III]/[N~II] (or its variant
([O~III]/H$\beta$)/([N~II]/H$\alpha$))\footnote{Since
H$\alpha$/H$\beta$~$\approx 2.86$ over a wide range of physical conditions,
[O~III] $\lambda 5007$/[N~II] $\lambda 6584$ and ([O~III]
$\lambda 5007$/H$\beta$)/([N~II] $\lambda 6584$/H$\alpha$)
carry essentially the same information. Here we will refer to the two
indicators interchangeably, but adopt the former in our own work.}
\citep[e.g.,][]{zahid+14, 
steidel+14, sanders+15}, and ([O~II] + [O~III])/H$\beta$, otherwise known as
the \citet{pagel+79} $R_{23}$ relation \citep[e.g.,][]{mannucci+10, belli+13,
nakajima+13, henry+13, cullen+14, maier+14}.  Each of these indicators has
significant limitations and/or systematic uncertainties.  For example,
abundances based on the (usually weak) [N~II] lines are susceptible to shifts
in the ionization parameter and N/O abundance ratio, metallicities that 
depend on [O~II]/[O~III] are sensitive to differential reddening, and the 
$R_{23}$ index is double-valued, while being only weakly dependent on 
metallicity when $12+\log{\rm (O/H)} \lesssim 8.4$ \citep[e.g.,][]{oey+00, 
kewley+02, kewley+08}. More importantly, all of these relations require the 
use of empirical calibrations that may not be applicable in the distant 
universe. To translate line ratios such as [O~III]/[O~II] or $R_{23}$ into 
metal abundances, one either needs to invoke radiative transfer models
\citep[e.g.,][]{kewley+02}, or compare to local galaxies whose electron
temperatures have been measured via the auroral [O~III] line at 4363~\AA{}
\citep[e.g.,][]{pettini+04, nagao+06, maiolino+08}. Unfortunately, these
procedures only work if the physical conditions that govern the strengths of
these prominent emission lines do not change with redshift.   This assumption
may not be true: several studies have demonstrated that at fixed stellar mass,
galaxies at $z \sim 2$ have higher ionization parameters
\citep[e.g.,][]{nakajima+14} and harder ionization fields
\citep{steidel+14,shapley+15} than their local universe counterparts, and, when
stellar mass and specific star formation rate (sSFR) are both held fixed, $z
\sim 3$ star-forming regions have electron densities that are a full order of 
magnitude higher than those seen at $z \sim 0.1$ \citep{shirazi+14}. 

Differences in physical conditions can produce offsets in the
strong-line ratio diagnostics, such as those used in the BPT diagram
\citep[e.g.,][]{bpt81, trump+13, kewley+13, juneau+14, coil+15}, and in
abundance determinations.    For example, in the $z \sim 0.1$ universe,
\citet{brinchmann+08} showed that young galaxies with large H$\beta$ rest-frame
equivalent widths (EW$_{{\rm H}\beta}$) have higher ionization parameters (due
to either higher electron densities or lower Lyman continuum optical depths),
and are offset from normal galaxies in the [O~III]/H$\beta$ versus
[N~II]/H$\alpha$ plane.  Similarly, when studying a local sample of luminous
($L_{{\rm H}\beta} > 3 \times 10^{40}$~ergs~s$^{-1}$) compact galaxies with
large H$\beta$ rest-frame equivalent widths (EW$_{{\rm H}\beta} > 50$~\AA),
\citet{izotov+11} found an offset between the oxygen abundances computed from
direct (weak-line) methods and the metallicities derived using strong-line
relations. With uncertainties such as these, it is no wonder that $z \sim 2$
studies of the MSZ have yielded conflicting results, with some groups finding
evidence for evolution \citep[e.g.,][]{zahid+14, steidel+14, cullen+14,
sanders+15}, while others detecting no evidence for change
\citep[e.g.,][]{mannucci+10, belli+13, henry+13, maier+14}.  

Here we address the question of evolution in the stellar
mass--star formation rate--metallicity plane using 256 emission-line galaxies
at $1.90 < z < 2.35$ derived from {\sl Hubble Space Telescope (HST)\/} grism
surveys of the COSMOS, GOODS-N, and GOODS-S fields \citep{brammer+12}. In this
redshift interval, the emission lines of [O~II] $\lambda 3727$, [Ne~III]
$\lambda 3869$, H$\gamma$, H$\beta$, and [O~III] $\lambda\lambda 4959, 5007$
all fall into the range of the near-infrared G141 grism, which allows for the
use of multiple strong-line indicators, such as [O~III]/[O~II],
[O~III]/H$\beta$, [O~II]/H$\beta$, [Ne~III]/[O~II], and $R_{23}$.  We calibrate
the strong-line relations following the guidelines of \citet{izotov+11}, and
carefully choose a local sample of Sloan Digital Sky Survey (SDSS) objects that
is matched in both SFR and stellar mass to our high-$z$ objects.   Stellar
masses are obtained from SED fitting, star formation rates from the UV
continuum, and extinction from the UV slope. We then compare these data to our
local sample, and to a simple theory proposed by \citet{dayal+13}.

In \S\ref{sec:sample}, we describe the dataset and the techniques used to
identify $1.90 < z < 2.35$ emission-line galaxies. In \S\ref{sec:SED} we detail
our stellar mass determinations, in \S\ref{sec:extinction} we discuss the issue
of nebular extinction and obscuration, in \S\ref{sec:SFR} we describe our
measurements of star formation rates, and in \S\ref{sec:linefluxes} we derive
emission line fluxes. In \S\ref{sec:metallicity} we describe our method for
deriving strong-line metallicities, both locally and at $z \sim 2.1$, using a
local sample of galaxies matched in stellar mass and star-formation rate. In
\S\ref{sec:fmr} we describe our results and show the relationship between
stellar mass, SFR, and metallicity of star forming galaxies at $z \sim 2.1$
galaxies is somewhat different from that seen locally, and we place our results
in the context of models.  Finally, in \S\ref{sec:summary}, we summarize our
results.

Throughout this paper, all quoted equivalent widths are in the rest-frame
unless otherwise stated, and we adopt a solar metallicity of $12+\log{\rm
(O/H)_{\odot}} = 8.69$ \citep{asplund+09}.  We also assume a flat $\Lambda$CDM
cosmology with $H_0 = 70$~km~s$^{-1}$~Mpc$^{-1}$, $\Omega_m = 0.3$, and
$\Omega_\Lambda = 0.7$ \citep{hinshaw+13, planck+14}.

%%%%%%%%%%%%%%%%%%%%%%%%%%%%%%%%%%%%%%%%%%%%%%%%%%%%%%%%%%%%%%%%%%%%%%%%%%%%%%%
\section{Sample Selection}
\label{sec:sample}
We chose for analysis three extragalactic fields with an abundance of
photometric and spectroscopic data:  COSMOS \citep{COSMOS}, GOODS-N, and
GOODS-S \citep{GOODS}.  In these regions, there exist deep optical and IR
images from the {\sl HST\/} CANDELS program \citep{CANDELS, koekemoer+11},
near-IR grism spectra from {\sl HST} \citep{brammer+12}, and supplemental
broad- and intermediate-bandpass photometry from a host of ground-based studies
\citep{skelton+14}.  By combining these data, we can measure the metallicities,
stellar masses, and star formation rates for a large sample of $z \sim 2.1$
galaxies spanning a wide range of parameter space.

To select our sample of $z \sim 2.1$ galaxies, we began with the G141 near-IR
grism data from the {\sl Hubble Space Telescope's\/} Wide Field Camera 3 (GO
programs 11600, 12177, and 12328; \citealp{kimble+08}).  This dataset, which is
the product of the 3D-HST \citep{brammer+12} and AGHAST \citep{weiner+14}
surveys, extends over 625~arcmin$^2$ and covers five well-studied fields,
including our targeted regions of COSMOS, GOODS-N, and GOODS-S\null.  When
combined with the surveys' direct F140W images and additional CANDELS F125W and
F160W photometry \citep{koekemoer+11}, these data provide $R = 130$ ($\sim
93$~\AA, 2-pixel FWHM) resolution slitless spectra between the wavelengths
$1.08~\mu$m and $1.68~\mu$m for all objects brighter than F140W $\sim 26$~mag. 
For galaxies in the redshift range $1.90 < z < 2.35$, this region covers many of
the important rest-frame optical emission lines, including [O~II] $\lambda
3727$, [O~III] $\lambda\lambda 4959,5007$, [Ne~III] $\lambda 3869$, H$\beta$,
and H$\gamma$.

A full description of the procedures used to reduce these grism frames and
identify our candidate galaxies is given by \citet{zeimann+14}.  Briefly, we
multi-drizzled the deep CANDELS images onto the reference frame defined by the
grism survey's shallower F140W images and created a master catalog of all
objects in the region brighter than F140W $\sim 26$~mag.
Using the {\tt aXe} software \citep{kummel+09}, we then
processed the spectral frames, subtracted a master sky frame from each image,
and extracted the 2-D spectrum of each object in the master catalog.  Sources
present on multiple frames were processed, drizzled to a common system
\citep{fruchter+09}, and co-added into a single higher signal-to-noise ratio
spectrum \citep{horne86}.  We then employed the optimal extraction algorithm
discussed by \citet{kummel+09} to create a 1-D spectrum for each object that
includes its flux density, the error on the flux density, and the contamination
fraction in units of flux density.  Finally, we created a web page containing
each object's F140W magnitude, ($x, y$) position, equatorial coordinates,
direct image cutout, grism image cutout, and 1-D extracted spectrum in both
counts and flux.  This process provided an easy and efficient way to review
each object, maintain quality control, and select galaxies for our analysis.

Since our investigation is limited to the redshift range $1.90 < z < 2.35$, we
visually examined each 1-D extracted spectrum, and searched for evidence of
emission from hydrogen (H$\beta$ and H$\gamma$), [O~II] $\lambda 3727$,
[Ne~III] $\lambda 3869$, and the distinctively shaped blended doublet [O~III]
$\lambda\lambda 4959,5007$.   Only galaxies with two or more emission lines
(i.e., a secure redshift) and a relatively clean spectrum (where contamination
from overlapping spectra is small) were chosen for analysis.  After excluding
four X-ray bright sources (i.e., likely AGN) and one object where our mass
determination did not converge, we obtained a final $z \sim 2.1$ sample of 
256 galaxies, mostly chosen via their bright [O~III] $\lambda 5007$
or [O~II] $\lambda 3727$ emission. 

\begin{figure}[t]
\ifpdf
    \centering
    \includegraphics[scale=0.75]{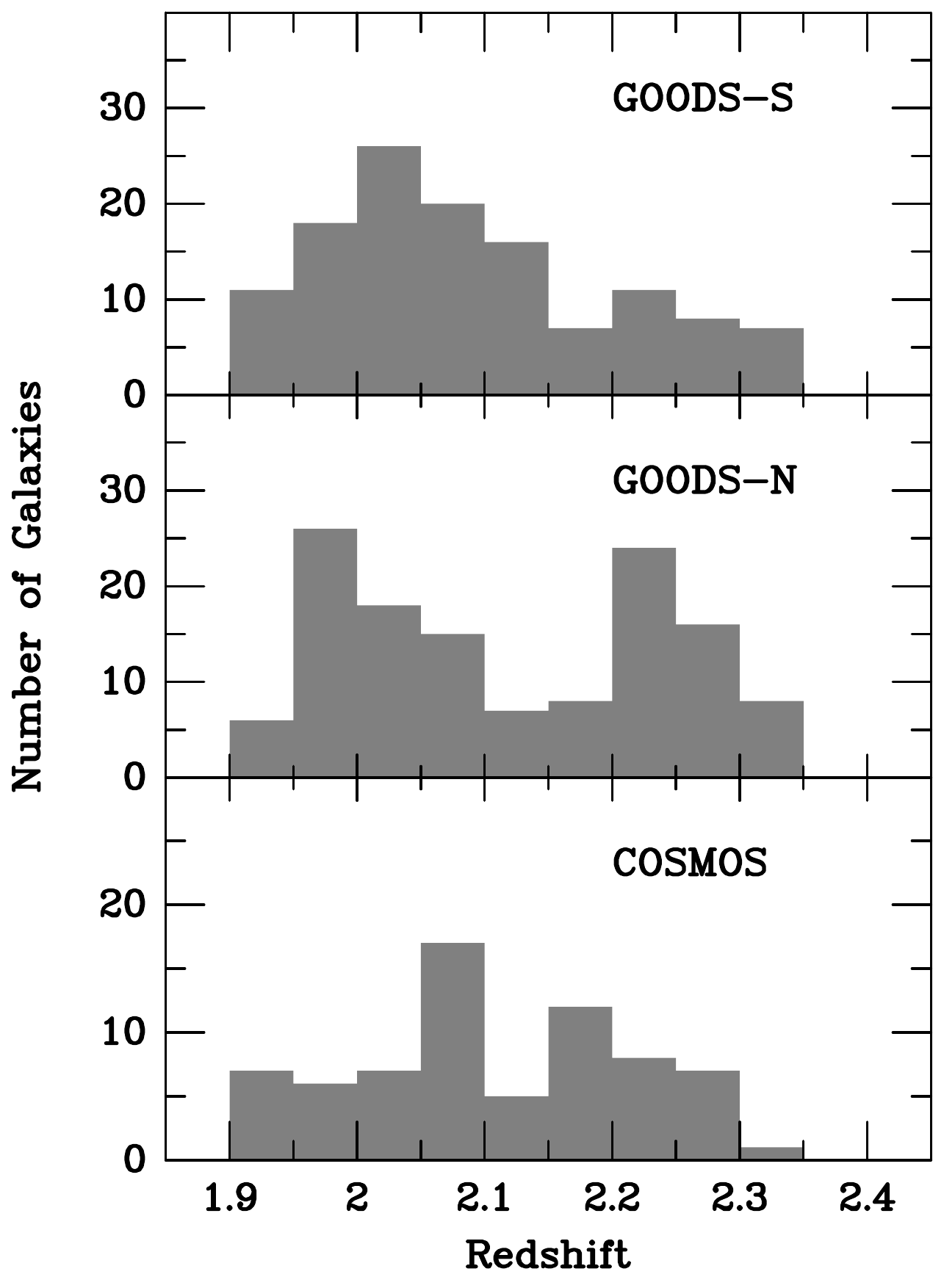}
\else
    \epsscale{0.5}
    \plotone{redshift.eps}
\fi
\caption{The redshift distribution of $1.90 < z < 2.35$ emission-line
galaxies in the COSMOS, GOODS-N, and GOODS-S fields.  The data show that the
sample contains galaxies from the full redshift range.}
\label{fig:redshift}
\end{figure}

The redshift range of these galaxies is shown in Figure~\ref{fig:redshift}. To
understand this distribution and the selection effects associated with this
sample, we created and ``observed'' a set of simulated emission-line spectra in
the exact same manner as our program data.  We began with the F140W magnitude
and positional distributions defined in the master catalog (see
\citealp{zeimann+14} for more details), and randomly assigned to each object a
redshift (between $1.90 < z < 2.35$), a metallicity ($7 < 12 + \log({\rm O/H})
< 9$), and a H$\beta$ flux (drawn from a uniform distribution in log space with
$-18 < \log (F_{{\rm H}\beta}) < -16$~ergs s$^{-1}$~cm$^{-2}$).   The relations
of \citet{maiolino+08} were used to translate these values into predicted line
strengths for [O~III], [Ne~III], [O~II], and H$\beta$, and the lines were then
superimposed onto a constant flux density continuum that matched the object's
F140W magnitude.  Our simulated spectrum was then placed onto a modeled grism
image (and an accompanying direct image) using the {\tt
aXeSim}\footnote{\url{http://axe.stsci.edu/axesim/}} software package. A total
of 500 of these simulated spectra were then ``observed'' and extracted in the 
exact same manner as the original data.  A summary of this analysis can be 
found in \citet{zeimann+14}. In brief, contamination from overlapping spectra 
excluded $\sim 15\%$ of the spatial area from our analysis. However, aside 
from this geometric factor, there is little variation in our ability to 
recover and measure spectra as a function of redshift, continuum luminosity, or
metallicity.   For the GOODS fields, our 50\% and 80\% line flux completeness
limits for the recovery of H$\beta$ are $\sim 10^{-17}$~ergs~s$^{-1}$~cm$^{-2}$
and $\sim 3 \times 10^{-17}$~ergs s$^{-1}$~cm$^{-2}$, respectively.  For the
COSMOS field, the recovery limits are shallower by a factor of $\sim 1.5$, due
to the higher sky background \citep{brammer+12}.

Our final catalog of 256 emission-line selected objects in the redshift
range $1.90 < z < 2.35$ is given in Table~\ref{tab:data}.

%%%%%%%%%%%%%%%%%%%%%%%%%%%%%%%%%%%%%%%%%%%%%%%%%%%%%%%%%%%%%%%%%%%%%%%%%%%%%%
\section{Stellar Mass Measurements}
\label{sec:SED}
The stellar masses for our $z \sim 2$ galaxies were computed by fitting
the objects' spectral energy distributions (SEDs) to stellar population models.
We began with the photometric catalog of \citet{skelton+14},
which combines the results of 30 distinct ground- and space-based imaging
programs into a homogeneous set of broad- and intermediate-band flux densities
in the wavelength range $0.35~\mu$m to $8.0~\mu$m.  In the COSMOS field, this
dataset contains photometry in 44 separate bandpasses, with measurements from
{\sl HST, Spitzer,} Subaru, and a host of smaller telescopes.  In GOODS-N, data
from {\sl HST, Spitzer,} Keck, Subaru, and the Mayall telescope are available
in 22 bandpasses, while in GOODS-S, six different telescopes provide flux
densities in 40 bandpasses.

To translate these photometric measurements into stellar mass and reddening, we
used the models of \citet{BC03}, which were updated in
2007 (BC07) with an improved treatment of the thermal-pulsing asymptotic giant
branch (TP-AGB) phase of stellar evolution.   (This update is not important for
the current work, since most of our galaxies are too young to have a
substantial TP-AGB population.)  For consistency with previous analyses of this
redshift era \citep[e.g.,][]{acq+11, hagen+14}, we adopted a \citet{kroupa01}
initial mass function (IMF) over the range $0.1 \, M_{\odot} < M < 100 \,
M_{\odot}$, and accounted for internal reddening via a \citet{calzetti01}
obscuration law.  Since stellar abundances are poorly constrained by broadband
SED measurements, we fixed the metallicity of our models to $Z = 0.2 \,
Z_{\odot}$, which is close to the median gas-phase abundance measured for our
sample (see below).  Emission lines and nebular continuum, which can be an
important contributor to the broadband SEDs of high-$z$ galaxies
\citep[e.g.,][]{schaerer+09, atek+11}, were modeled following the prescription
of \citet{acq+11} with updated templates from \cite{acq+12}.   We also excluded
from consideration all bandpasses redward of (rest-frame) $3.3~\mu$m, where
interstellar PAH features may contribute to the continuum flux density
\citep{tielens08}, and blueward of Ly$\alpha$, where the statistical correction
for intervening Ly$\alpha$ absorption \citep{madau95} may not always be
appropriate.

Finally, we assumed that the SFRs of our galaxies have been constant with time.
Obviously, this last premise is a simplification:  our $z \sim 2.1$ galaxies
can have any number of star formation histories, and in fact, several recent
studies \citep[][and references therein]{reddy+12, pacifici+13, salmon+15} have
argued that constant or declining star formation rates are not consistent
with $z \sim 2$ data.  However, at the epochs under
consideration (only $\sim 3$~Gyr after the Big Bang), the exact history of star
formation has little bearing on our results.  Indeed, \citet{reddy+12} has
shown that stellar masses derived using the constant star formation rate
assumption are essentially identical to those computed using a $\tau$-model
with an increasing rate of star formation (see their Figure~8).   In the case
of our grism-selected systems, this result is confirmed: the use of an
exponentially increasing SFR with an e-folding time of 100~Myr would decrease
our stellar masses by only $0.002 \pm 0.097$~dex, and our best-fit stellar
reddenings would increase by just $\Delta E(B-V) = 0.007 \pm 0.027$.

Since SED fitting is a notoriously non-linear problem that may involve many
local minima, highly non-Gaussian errors, and degeneracies between parameters,
we chose to analyze our data using {\tt GalMC}, a Markov-Chain Monte-Carlo
(MCMC) code with a Metropolis-Hastings sampler \citep{acq+11}. Using four
chains with random starting locations, we fit each SED with three free
parameters: stellar mass, $E(B-V)$, and age.  (For a constant SFR history, age
and mass are uniquely related to star formation rate.)  Once completed, the
chains were analyzed via the {\tt CosmoMC} program {\tt GetDist}
\citep{lewis+02}, and, since multiple chains were computed for each object, the
\citet{gelman+92} $R$ statistic was used to test for convergence via the
criterion $R - 1 < 0.1$ \citep{brooks+98}.

In general, the stellar masses associated with our fits are reasonably robust,
with typical statistical uncertainties of $\sim 0.3$~dex.  Of course, there are
additional systematic errors associated with our results. For example, the use
of a \citet{chabrier03} IMF as opposed to a \citet{kroupa01} IMF 
would systematically reduce our stellar mass estimates by a few tenths of a 
dex, while the assumption of solar metallicity could change our masses by $\sim
0.1$~dex.   In theory, a change from the BC07 models to the population
synthesis models of 2003 could also make a difference, though the analysis of
\citet{hagen+14} showed that in the case of our $z \sim 2.1$ systems, this 
offset is negligible.  A full discussion of these systematics is given by
\citet{conroy13}.

Table~\ref{tab:properties} summarizes the properties of all our $1.90 < z <
2.35$ 3D-HST galaxies.  As the top left panel of 
Figure~\ref{fig:histograms} shows, our sample of grism-selected
star-forming galaxies have a wide range of stellar mass, extending
from $\sim 10^{10} M_{\odot}$ to values as low as $\sim 10^7 \, M_{\odot}$.  A
local SDSS galaxy sample is also shown in the figure; these comparison
data, along with the other panels of the figure, will be described below.

%%%%%%%%%%%%%%%%%%%%%%%%%%%%%%%%%%%%%%%%%%%%%%%%%%%%%%%%%%%%%%%%%%%%%%%%%%%%%%
\section{The UV Continuum and Nebular Extinction}
\label{sec:extinction}
The wavelength range of our grism spectroscopy includes all the rest-frame
optical emission lines blueward of $\sim 5100$~\AA, so, in theory, we could
obtain an estimate of nebular reddening directly from the Balmer decrement.  In
practice, however, H$\gamma$ and the higher-order Balmer lines are generally
too weak (and too close in wavelength to H$\beta$) to produce a reliable
measure of extinction.  Thus, in the absence of H$\alpha$ measurements, 
we must rely on the stellar spectral energy distribution to constrain the
nebular extinction.

Although our SED fits do produce a measure of stellar extinction, these values
are subject to the multitude of assumptions associated with full spectrum SED
modeling. A much more direct path to the stellar reddening, which is more
reproducible by the reader and less subject to catastrophic errors, is to
simply use the slope of the UV continuum between 1250~\AA\ and 2600~\AA\null.
Because all of our $z \sim 2.1$ galaxies are undergoing vigorous star 
formation, their slope in this spectral range is well-approximated by a power 
law, i.e., 
\begin{equation}
F(\lambda) \propto \lambda^{\beta_0}
\label{eq:powerlaw}
\end{equation}
where $\beta_0 = -2.25$ for populations which have been forming stars for more
than $\sim 0.5$~Gyr, and is only slightly steeper for younger systems
\citep{calzetti01}.  We therefore adopt $\beta_0 = -2.25$ for our analysis; if
the UV slope is observed to be flatter than this, then the most-likely
explanation is obscuration due to dust.

Translating the observed power-law slope into a measure of total (stellar)
extinction at 1600~\AA\ and total (nebular) reddening requires a number of
assumptions. The most common assumption is that based on the work of
\citet{calzetti01}, who showed that in the starbursting galaxies of the local
universe, the observed slope of the rest-frame UV continuum, $\beta$, is
related to the total stellar extinction at 1600~\AA\ by 
\begin{equation}
A_{1600} = \kappa_{\beta} \left( \beta - \beta_0 \right) 
\label{eq:1600}
\end{equation}
with $\kappa_{\beta} = 2.31$.  \citet{calzetti01} also showed that for nebular
emission lines, a screen-model reddening law, such as that detailed by
\citet{cardelli+89} is appropriate, and that the amount of extinction 
affecting the gas is greater than that for the stars, so 
\begin{equation}
E(B-V)_{\rm stars} = 0.44 E(B-V)_{\rm gas} \quad {\rm and} \quad
A_{{\rm H}\beta} = \zeta_{{\rm H}\beta}  A_{1600} 
\label{eq:ex-hbeta}
\end{equation}
where $\zeta_{{\rm H}\beta} = 0.83$.  Whether this behavior continues at high
redshift is uncertain: while a number of surveys have presented evidence for
the applicability of this law at $z \sim 2$ \citep{forster+09, mannucci+09,
wuyts+13, price+14, holden+14, zeimann+14}, counter examples do exist
\citep[e.g.,][]{erb+06b, kashino+13}.

For consistency with other high-$z$ studies, we use the \citet{calzetti01}
obscuration law for our analysis.  For the emission-line selected galaxies
considered here, this law implies a median nebular extinction of $E(B-V)_{\rm
gas} = 0.21$\null. Note that if we were to assume that the \citet{calzetti01}
law systematically under-predicts the color excess of the gas by $E(B-V)_{\rm
gas} \sim 0.1$, our median galactic oxygen abundance would increase by only
0.14~dex (see below). Similarly, if instead of applying an individual
extinction correction to each galaxy, we were to adopt $E(B-V)_{\rm gas} = 0.2$
for the entire sample, or used the \citet{garn-best10} correlation between
extinction and stellar mass, our results would be qualitatively similar.
Finally, the $E(B-V)$ obtained from the UV slope is systematically
lower by $\sim0.13$~dex than that obtained from SED fitting. Once again, this
difference is small enough that it does not affect our conclusion.

%%%%%%%%%%%%%%%%%%%%%%%%%%%%%%%%%%%%%%%%%%%%%%%%%%%%%%%%%%%%%%%%%%%%%%%%%%%%%
\section{Galactic Star-Formation Rates}
\label{sec:SFR}
Our SED analysis also produces a measure of star formation rate that is
time-averaged over the history of the galaxy.  To obtain an estimate of a
system's present-day SFR, we have two choices: we can use our grism
measurements of the H$\beta$ line, which is produced by the recombination and 
cascade of electrons photo-ionized by hot $M \gtrsim 10 \, M_{\odot}$ stars, 
or we can rely on the photometry of the rest-frame UV continuum, which traces 
the starlight from objects with $M \gtrsim 5 \, M_{\odot}$.  The advantage of 
the former method is that H$\beta$ records an ``instantaneous'' SFR, since 
only stars with ages less than $\sim 10$~Myr contribute to its emission 
\citep{kennicutt+12}.  A major disadvantage, however, is that H$\beta$ is 
rather weak in our spectra, and its measurement can be affected by underlying 
stellar absorption.  (\citealt{zeimann+14} has shown that this equivalent 
width correction is likely to be small in our grism-selected galaxies, but it 
is still present.)  Moreover, although the method is well-calibrated in the 
local universe \citep{hao+11, murphy+11, kennicutt+12}, it is much more 
sensitive to metallicity than some other techniques, and this can lead to 
factor of $\sim 2$ errors at $z \sim 2.1$ \citep{zeimann+14}.  Finally, 
without knowledge of the Balmer decrement, SFR corrections for internal 
extinction can be problematic, as it is well known that the dust column
that affects galactic emission lines is systematically greater than that
which reddens starlight \citep[e.g.,][]{charlot+00, calzetti01}.  Since our
measurement of extinction is limited to the slope of the rest-frame UV
continuum, our knowledge of intrinsic H$\beta$ luminosities is indirect at
best.

The alternative approach is to use the brightness of the rest-frame UV
continuum as a measure of star formation rate.  UV measurements are more
sensitive to extinction as those of H$\beta$ (see equation~\ref{eq:ex-hbeta}),
and the method traces star formation over a timescale that is $\sim 10$~times
longer \citep[e.g.,][]{kennicutt+12} than that traced by the Balmer
lines.  However, its calibration is much less
sensitive to metallicity variations than that for emission-line SFR indicators
\citep{zeimann+14}, and in most cases, the UV photometry of \citet{skelton+14}
has a higher signal-to-noise than our grism spectroscopy (H$\beta$ has a
S/N~$<2$ on average).  We therefore adopt the de-reddened continuum flux
density at rest-frame 1600~\AA\ as our primary SFR indicator, with the
calibration given by \citet{kennicutt+12} (originally tabulated in
\citealp{murphy+11} and \citealp{hao+11}) as our transformation coefficient.
Figure~\ref{fig:histograms} shows the distribution of these star formation
rates.

Since most local universe SFR measurements are based on Balmer-line emission,
and since the SFRs tabulated in the Sloan Digital Sky Survey (SDSS) Data
Release~7 MPA-JHU galaxy
catalog\footnote{\url{http://www.mpa-garching.mpg.de/SDSS/DR7}} are based on
H$\alpha$, we will examine if this choice of SFR indicator affects our
conclusions about the $z \sim 2.1$ MSZ relation.

%%%%%%%%%%%%%%%%%%%%%%%%%%%%%%%%%%%%%%%%%%%%%%%%%%%%%%%%%%%%%%%%%
\section{Emission Line Fluxes}
\label{sec:linefluxes}
\begin{figure}[t]
\centering
\includegraphics[scale=0.95]{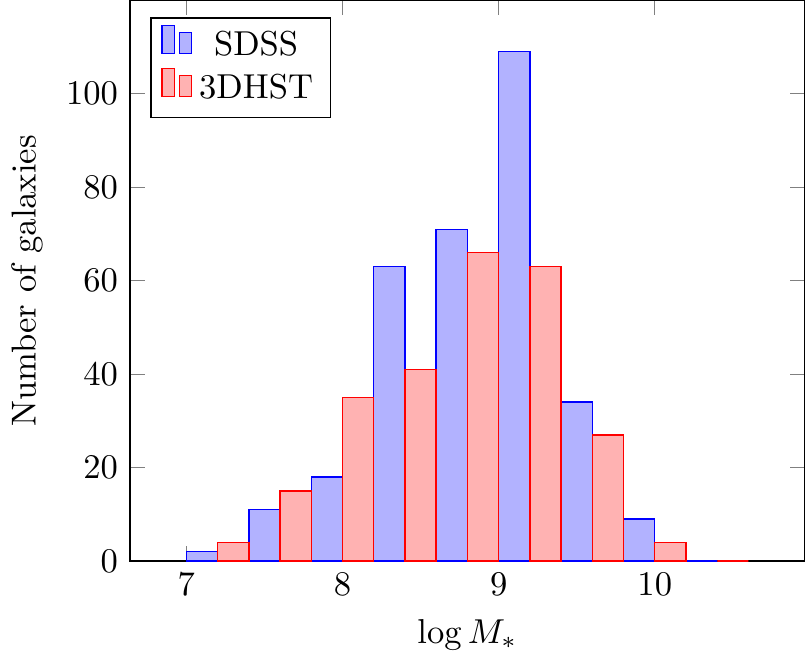}
\includegraphics[scale=0.95]{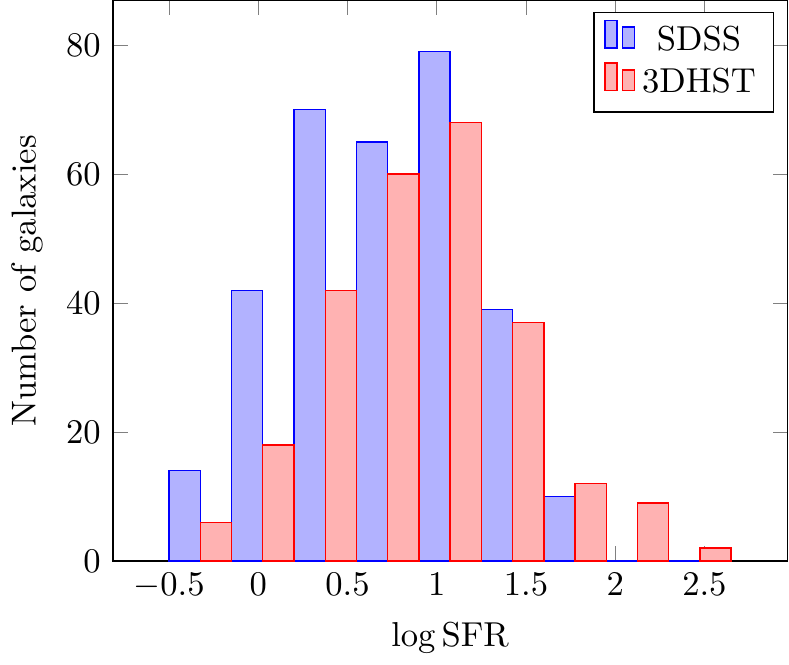}
\includegraphics[scale=0.95]{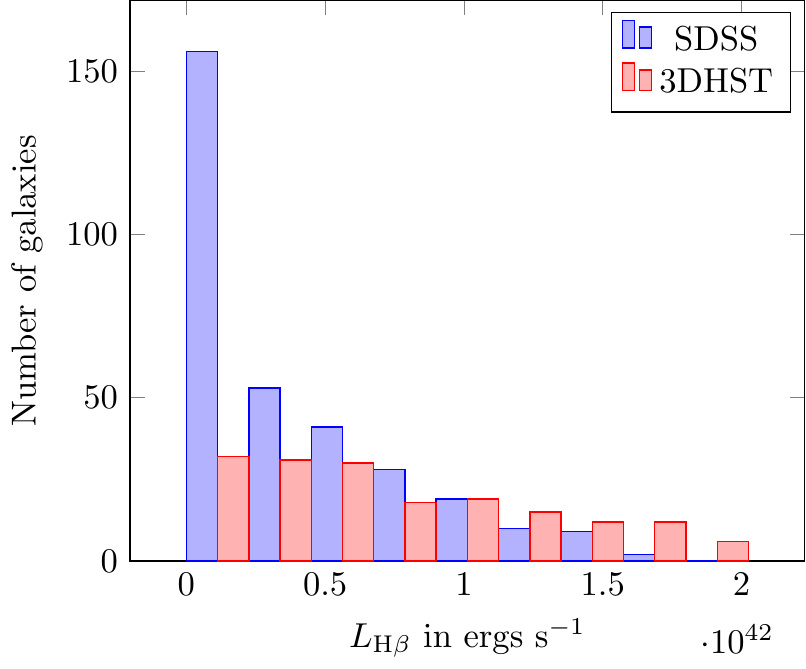}
\includegraphics[scale=0.95]{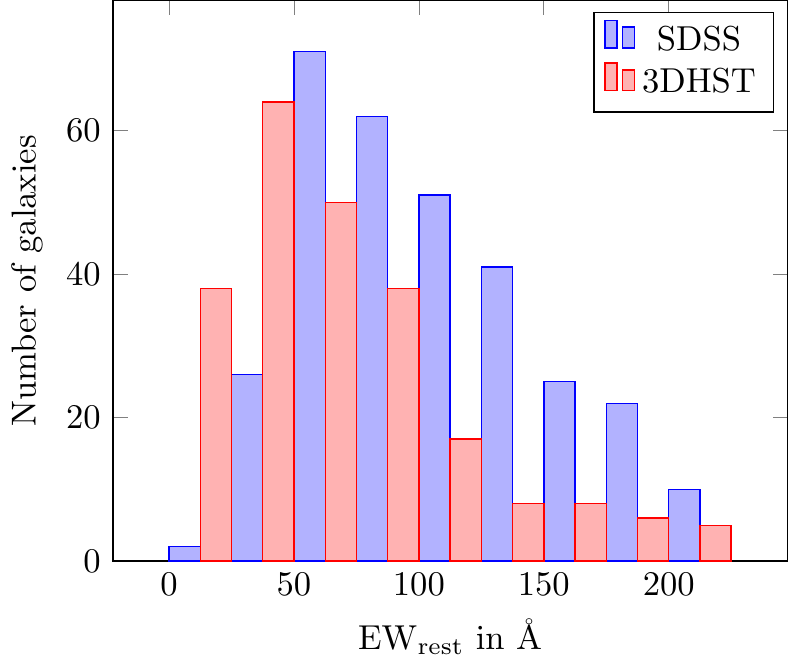}
\caption{Histograms of the stellar mass, star formation rate, H$\beta$
luminosity, and rest-frame H$\beta$ equivalent width for the {\sl HST-}selected
star-forming galaxies at $z \sim 2.1$ and a set of local SDSS systems matched
in stellar mass and star formation rate.  Although the two distributions are
different in detail, they both cover similar regions in mass--SFR--emission 
line phase space.}
\label{fig:histograms}
\end{figure}

For the redshift range under consideration, the {\sl HST\/} grism spectra
contain six strong emission lines:  [O~II] $\lambda 3727$, [Ne~III] $\lambda
3869$, H$\gamma$, H$\beta$, and [O~III] $\lambda\lambda 4959,5007$.  To measure
their fluxes, we began with the assumption that each spectrum could be fit via
a series of Gaussian-shaped emission line profiles superimposed on a
fourth-order ($n = 4$) polynomial continuum.  In other words, we assumed the
intrinsic spectrum could be modeled via
\begin{equation}
    f(\lambda) = \sum_{i=0}^n c_i \lambda^i + \sum_{j=n+1}^{n+k} \frac{c_j}{\sqrt{2 \pi}
\sigma} \exp \left( - \frac{(\lambda - \lambda_{e,j})^2}{2 \sigma^2} \right) +
w(\lambda)
\label{eq:model}
\end{equation}
where $w(\lambda)$ represents Gaussian white noise at each wavelength, the
values $\lambda_e = (1 + z) \,  \lambda_0$ are the observed wavelengths of the
$k = 6$ fitted lines, and the values of $c_j$ encode the strengths of those 
lines.  Following \citet{storey+00}, the ratio of the [O~III]
doublet was fixed at 2.98:1.  We then considered the fact that the data
generated by the G141 grism are undersampled, so that each point in the
measured spectrum, $F(x)$, is actually the integral of the model function over
the size of the pixel $\Delta x$, i.e.,
\begin{equation}
F(x) = \int_{x - \Delta x / 2}^{x + \Delta x/2} \, f(\lambda) d\lambda
\label{eq:pixelation}
\end{equation}
We fit the observed spectrum to this model $F(\vec x)$ via
the maximum likelihood method.  A full description of this procedure is given
in the Appendix, and sample fits are shown in Figure~\ref{fig:spectra}.
Finally, to correct for stellar absorption we augmented the H$\beta$
fluxes using the Balmer absorption line equivalent widths derived from our 
best-fit model SEDs.  The derived line fluxes for
all the $1.90 < z < 2.35$ galaxies are presented in Table~\ref{tab:data}, and a
histogram of H$\beta$ luminosities is displayed in Figure~\ref{fig:histograms}.

\changed{

The galaxies in our program are generally very faint, with $22 \lesssim m_H
\lesssim 26$, hence the signal-to-noise in the grism data is often quite low.
However, as described in \S\ref{sec:sample}, our galaxies were primarily
selected via the presence of the distinctively shaped blended doublet
[O~III]~$\lambda\lambda 4959, 5007$.  This fixes the redshift of each object,
and allows us to measure lines such as H$\beta$, [O~II] $\lambda 3727$, and
[Ne~III] $\lambda 3869$ to a very low signal-to-noise. Histograms of the ratio
of the measured line fluxes $c_i$ to their formal statistical uncertainties
$\delta c_i$ are shown in Figure~\ref{fig:snr}. Note that $c_i$ is not the true
value of the emission line flux, nor is $\delta c_i$ the uncertainty on this
value.  Rather, they are the observed realizations of the true signal $S$ and
the noise in the measurement, $N$\null.   In the limit where $S/N \gg 1$, the
ratio $c_i/\delta c_i \rightarrow S/N$,\footnote{This can be seen by
recognizing that $c_i \sim S \pm \delta c_i$.} but for faint sources, the
stochasticity will dominate, resulting in some objects having values of $c_i <
0$. Our Bayesian approach to metallicity determination properly accounts for
these low S/N measurements and the resultant probability distribution functions
for metallicity are propagated throughout our analysis (see
\S\ref{sec:metallicity}).

}

Our $z \sim 2.1$ galaxies are all undergoing vigorous stellar formation.
The best way to illustrate this fact is through the distribution of 
H$\beta$ emission-line equivalent widths.   For the vast majority of our 
$z \sim 2.1$ objects, the {\sl HST\/} grism spectra do not go deep enough to 
reliably detect the stellar continuum.  Nevertheless,
we can compare the strengths of our observed emission-line fluxes to the 
objects' stellar continua via the {\sl HST's\/} broadband infrared photometry.
We began by taking each galaxy's F140W magnitude, and subtracting the 
contribution of all the emission lines falling into its bandpass.  We then 
scaled the emission lines to our photometrically-estimated continuum 
measurement, i.e., 
\begin{equation}
{\rm EW}_{{\rm H}\beta} = \frac{c_{{\rm H}\beta}}{f_{\rm F140W} \Delta\lambda 
- \sum_i c_i} \, \Delta\lambda
\end{equation}
where the bandwidth of the F140W filter is $\Delta\lambda=3840$~\AA{}, the
values of $c_i$ represent the monochromatic fluxes of the emission lines
falling within the filter (see the appendix), and $f_{\rm F140W}$ is the
system's $H$-band flux density.  Each observed equivalent width was then 
divided by $1+z$ to obtain its value in the rest frame.  These values are 
displayed (along with their uncertainties) in Figure~3 of \citet{zeimann+14}.
To examine the true distribution of H$\beta$ emission-line
equivalent widths, we correct our values for the Balmer absorption of the
stars using our best-fit model for the SED. Finally, we also correct for the
difference between stellar and nebular reddening via the \citet{calzetti01}
relation $E(B-V)_{\rm stellar} = 0.44 E(B-V)_{\rm gas}$.

Obviously, equivalent widths derived from a combination of spectroscopy and
broadband photometry can carry a significant uncertainty.  
In particular, for many of our program galaxies, the contribution of 
[O~III] $\lambda 5007$ to the F140W flux density is substantial, so for 
the faintest objects, the H$\beta$ line's equivalent width is poorly known
at best.  In fact, it is these objects which dominate the low-EW end of the 
histogram displayed in Figure~\ref{fig:histograms}.  All the well-measured 
galaxies in our sample have H$\beta$ equivalent widths greater than 30~\AA, 
and most have EW$_{{\rm H}\beta} > 50$~\AA\null.  The objects detected by
the G141 grism are clearly dominated by on-going star formation.

%%%%%%%%%%%%%%%%%%%%%%%%%%%%%%%%%%%%%%%%%%%%%%%%%%%%%%%%%%%%%%%%%%%%%%%%%%%%%%
\section{Deriving Gas-Phase Metallicities using Local Counterparts}
\label{sec:metallicity}

Our $z \sim 2.1$ grism observations do not reach the weak nebular lines of
[O~III] $\lambda 4363$ and He~II $\lambda 4686$. Consequently,
the only way to translate our bright-line emission fluxes into gas-phase
metallicities is to do so differentially, using systems with well measured
oxygen abundances as a control sample.  By examining the behavior of the bright
emission lines as a function of nebular abundance, one can define a metallicity
calibration that can be applied to the galaxies of the distant universe.

Unfortunately, such a procedure is fraught with difficulties.  Perhaps the most
commonly used set of strong-line metallicity calibrations are those derived by
\citet{maiolino+08}, who combined direct abundance measurements for $\sim 300$
low-metallicity galaxies \citep[see][]{nagao+06} with photoionization-model
based abundances for $\sim 22,000$ high-metallicity systems \citep{kewley+02}.
However, the applicability of these relations to our study is unclear, as the
physical conditions within most nearby galaxies are significantly
different from those found in the high-redshift universe.  On average, $z \sim
2$ galaxies form stars ten times more rapidly than do local objects
\citep{whitaker+14}, and as discussed earlier, they tend to have harder
ionizing fields \citep{steidel+14}, higher ionization parameters
\citep{nakajima+14}, and higher nebular densities \citep{shirazi+14} for the
same stellar mass.    Consequently, the use of abundance calibrations that are
based on magnitude-limited samples defined from ensembles of low-$z$
galaxies may not be appropriate.  This is especially true at the
poorly-populated, low-luminosity, metal-poor end of the distribution 
($12+\log({\rm O}/{\rm H})<7.7$), where the \citet{maiolino+08} sample 
contains just a small handful of objects.

\subsection{The Local Stellar Mass- and SFR-Matched Sample}
\label{sec:matches}

Rather than use the entire SDSS catalog, which includes galaxies with star
formation rates and masses that are vastly different from those found in our $z
\sim 2.1$ galaxy sample, we chose to follow the example of \citet{izotov+06}
and compare to a set of galaxies whose global properties are similar to those
of our $z \sim 2.1$ systems.  To define this sample, we began with the table 
of emission line fluxes, stellar masses, and star formation rates provided by 
the Sloan Digital Sky Survey (SDSS) Data Release 7 MPA-JHU galaxy 
catalog\footnote{\url{http://www.mpa-garching.mpg.de/SDSS/DR7}}.  We
then selected those galaxies with H$\beta$ luminosities greater
than $L_{{\rm H}\beta} \geq 3\times10^{40}$~ergs~s$^{-1}$, and positive
monochromatic emission line fluxes for [O~II] $\lambda 3727$, 
[Ne~III] $\lambda 3869$, H$\beta$, and [O~III] $\lambda 5007$.  After
excluding objects with unusually strong [O~III] $\lambda 4363$ or
[N~II] $\lambda 6584$, or line ratios that suggested the presence of an 
AGN \citep{kauffmann+03}, we went through each $z \sim 2.1$ galaxy in our sample
and identified the five SDSS galaxies closest to it in $\log M_*$-$\log$~SFR
space.  (In the median, these ``matched'' galaxies were all within 0.12~dex of 
their $z \sim 2.1$ counterpart.)  Since some SDSS galaxies are associated with 
more than one $z \sim 2.1$ system, the resulting ``matched'' dataset 
consisted of 319 local galaxies.

Necessarily, this procedure of identifying counterparts to our $z \sim 2.1$
grism-selected galaxies is not ideal.  Unlike our COSMOS and GOODS galaxies,
which have comprehensive rest-frame UV through rest-frame IR photometry in up
to 44 separate bandpasses (see \S\ref{sec:SED}), the photometric data for the
local objects are limited to the 5 SDSS bandpasses, $ugriz$.  Consequently, the
cataloged SED-based masses are not as robust as those for the distant galaxies.
Similarly, the SFR estimates for the SDSS galaxies are based on H$\alpha$
\citep{brinchmann+04}, while those for the $z \sim 2.1$ objects rely on
photometry of the rest-frame UV\null.  While both techniques are
well-calibrated in the local universe \cite[e.g.,][]{kennicutt+12}, their
effective timescales are different (see \S\ref{sec:SFR}).  Finally, it is
well-known that galaxies at high redshift generally have systematically higher
SFRs than local systems \citep[e.g.][]{madau+14,whitaker+14}, and, despite our
best attempts to minimize the discrepancy, this causes our $z \sim 2.1$ systems
to extend over a wider range in stellar mass-SFR space than the local matched
sample.  As we will show below, we can address this problem by restricting our
comparison to galaxies with the same mass-SFR bin.  Nevertheless, it is a
fundamental limitation of the matched sample.  

The propriety of local mass- and SFR-matched sample of high line-luminosity
galaxies is illustrated in Figure~\ref{fig:lineratios}.  From the figure, 
it is clear that the envelope defined by our local matched sample and the
set of $z \sim 2.1$ grism-selected galaxies extend over the exact same region 
in emission-line ratio phase space ([O~III]/[O~II] vs.\ $R_{23}$).  Moreover, 
at high metallicity, the line ratios of both galaxy samples follow the 
track defined by the commonly-used metallicity calibration of 
\citet{maiolino+08}.  However, in the lowest metallicity systems 
($12+\log {\rm (O/H)} < 7.7$), the \citet{maiolino+08} relations are not a 
good match to the data, as they extend into a region of phase space that is 
not populated by either dataset.   Clearly, in the low-metallicity regime, a 
more robust calibration is needed.

\subsection{Strong Line Abundance Calibrations from the Local Sample}
\label{sec:calibration}

All but 47 of the galaxies in the local sample have measurements of the auroral 
[O~III] $\lambda 4363$ line that are at least twice that of its flux error.  
For these 272 objects, we can determine electron temperatures directly,
and hence derive a physics-based estimate of their gas-phase metallicity
\citep{AGN3, izotov+06}.  Using these data, we could obtain a new set of 
polynomial relations relating the galaxies' bright emission-line
ratios to these abundances.

To do this, we began by assuming that, due to the range of possible
physical conditions within bright H~II regions, there is no unique relation
between line ratio and metallicity.  Instead, we assumed that a system with
metallicity $Z$ can have a range of line ratios, with the probability of
observing any given line-ratio, $R$, defined as a Gaussian, with a mean $\log
R_i(Z) = r_i(Z)$ and a dispersion $\sigma_{r_i}$.  Here, $r_i(Z)$ is a
low-order polynomial, representing the behavior of the index versus
metallicity. In other words, the probability of observing a line ratio,
$R_i$, in a galaxy whose true metallicity is $Z$ is given by the Gaussian
distribution function\footnote{An alternative notation is to say
that the random variable $\log R_i$ is distributed as a Gaussian with mean
$r_i(Z)$ and variance $\sigma_{r_i}^2$, e.g. $\log R_i \sim
N(r_i(Z),\sigma_{r_i}^2)$.} $p(\log R_i) =
(2\pi\sigma_{r_i}^2)^{(-1/2)}\,\exp(-(\log R_i - r_i(Z))^2/2\sigma_{r_i}^2)$.
We then used the method of maximum likelihood to solve for the polynomial
coefficients of $r_i(Z)$ and $\sigma_{r_i}$ for the various line ratios
associated with the forbidden lines of oxygen, neon, and nitrogen. The
coefficients for these fits, along with the most-likely value of
$\sigma_{r_i}$, are given in Table~\ref{tab:relations} and displayed in 
Figure~\ref{fig:newmaiolino}.  Also shown in Figure~\ref{fig:newmaiolino}
are the more established relations of \citet{maiolino+08}.  
As expected, due to the physics of nebular
cooling and stellar opacities, the fits for the [O~III]/H$\beta$,
[O~II]/H$\beta$, and the $R_{23}$ relation
required a quadratic term \citep[e.g.,][]{kewley+02, kewley+13,
levesque+14}; for the other line ratios, linear fits were sufficient.

It is important to note that for our sample of mass-SFR matched,
high line-luminosity objects, our new relations are quite similar to
those of \citet{maiolino+08}, except in the regime where the oxygen abundance
dips below $12+\log({\rm O/H}) < 7.8$.  At these very
low metallicities, the \citet{maiolino+08} relations are defined by
literature measurements of roughly two dozen local, heterogeneously selected
galaxies \citep[see][]{nagao+06}.  In terms of numbers, our local sample is no
better, with only $\sim 4$ objects in this low-metallicity range.
However, because we selected our local galaxies to match the $z \sim 2.1$
systems' star formation rates and stellar masses,
it is likely that the physical conditions in these systems are a better
match to the high redshift counterparts.  We therefore adopted these new
calibrations for both the $z \sim 2.1$ galaxies and our local comparison
sample, and use them in our analysis.

\subsection{Deriving Metallicities}
\label{sec:meta-indicator}
To translate our emission line fluxes into galactic metallicities, we began by
choosing three strong-line abundance indicators for analysis:  $R_{23}$,
[O~III]/[O~II], and [Ne~III]/[O~II].  (Other combinations of indicators are
possible, but as long as the degrees of freedom are reduced to the minimum, the
results of each set of indicators are consistent.) We used the re-calibrated
versions of these relations discussed in \S\ref{sec:calibration} to construct
a ``meta-indicator'', that, given a metallicity and extinction, simultaneously
predicts the relative values of all the strong-line fluxes.  We then adopted a
nebular extinction law \citep{cardelli+89}, used the slope of the UV continuum
to obtain an estimate of the total extinction at H$\beta$ (see
\S\ref{sec:extinction}), corrected our line fluxes for this reddening while
fixing the H$\gamma$/H$\beta$ ratio to 0.47 \citep{hummer+87}, and applied the
metallicity calibrations shown in Figure~\ref{fig:newmaiolino}.  Finally,  we
computed the likelihood that any given galaxy with oxygen abundance $12 +
\log({\rm O/H})$ would possess the line fluxes given in Table~\ref{tab:data}.

The details of this procedure are explained in the Appendix. The constraining
power added by each indicator is illustrated in Figure~\ref{fig:metal_demo},
where we show the metallicity solution for one of the galaxies in our sample.
In the figure, the solid black horizontal lines and their surrounding dashed
lines represent the de-reddened values for $R_{23}$, [O~III]/[O~II], and
[Ne~III]/[O~II], and their uncertainties.  For comparison, the blue curves
display the metallicity calibrations  shown in Figure~\ref{fig:newmaiolino}.
Because in the construction of the likelihood function, the predicted variance
depends on both metallicity and line flux (see Appendix), the most likely
metallicity is not necessarily at the exact location where the black line and
blue curves intersect (though it is close).   The lower section of each plot
displays the likelihood that the observed line ratio is the product of a given
metallicity. These likelihoods were normalized in the metallicity range 7--9,
and are reproduced graphically in the lower-right hand panel of the
figure, along with the likelihood derived from the combined meta-indicator.  
For this particular object, it is apparent that the $R_{23}$ index does not
provide a strong constraint on metallicity, as the probability density
distribution function is broad and slightly double-peaked.  The [O~III]/[O~II]
ratio helps constrain this peak, as does (to a lesser extent) the ratio of
[Ne~III] relative to [O~II]\null.   

To test our maximum-likelihood procedure, we applied our code to the emission
lines of the local calibration galaxies.   The results are displayed in
Figure~\ref{fig:test_stats}, where we compare the metallicities computed from
our meta-indicator against those derived via ``direct'' analyses using the
temperature sensitive auroral [O~III] $\lambda 4363$ line. Although there may
be a slight trend in the data, a maximum likelihood analysis assuming constant 
scatter yields a slope of $0.998 \pm 0.050$, and, on average, the offset 
between the two abundances is $\log({\rm O/H})_{\rm strong-line} - 
\log({\rm O/H})_{\rm direct} = 0.028 \pm 0.154$~dex. This is significantly 
better than the results which would be obtained using the \citet{maiolino+08} 
relations ($-0.041 \pm 0.148$~dex), though the scatter about the one-to-one 
line is slightly higher.  The fact that there is a very small, 
metallicity-dependent trend in the residuals is due to our use of a 
first-order polynomial for the [O~III]/[O~II] component of the measurement.  
A second-order polynomial fit would remove this trend, but it would also cause 
our meta-indicator to become less reliable for low signal-to-noise objects.

Figure~\ref{fig:spectra} illustrates the dependence of our abundance estimates
on nebular extinction.  The most obvious feature to note is the lack of direct
constraints on reddening.  In many cases, this leads to double-valued
solutions:  a metal-poor solution, where the relative weaknesses of [O~II]
$\lambda 3727$ and [Ne~III] $\lambda 3869$ are presumed to be intrinsic, and a
metal-rich ($\sim$ solar metallicity) solution, where the short wavelength
lines are assumed to be depressed by extinction.  Conversely, for many galaxies
the oxygen abundance is reasonably well defined even without a reddening
constraint, with typical fitting errors of $\sim 0.8$~dex.  The application of
a stellar continuum-based reddening constraint \citep{calzetti01} 
then reduces this uncertainty to $\sim 0.4$~dex.

%%%%%%%%%%%%%%%%%%%%%%%%%%%%%%%%%%%%%%%%%%%%%%%%%%%%%%%%%%%%%%%%%%%%%%%%%%
\section{The (Non-) Fundamental Mass--Metallicity--SFR Relation}
\label{sec:fmr}

In \S\ref{sec:intro}, we discussed the numerous investigations aimed at
measuring the relationship between stellar mass, star formation rate, and
metallicity in the $z \sim 2$ universe.  At the same time, there has been a
similarly large effort concerning the Fundamental Metallicity Relation of the
local universe, with various authors choosing to emphasize differences in their
sample selection, stellar-mass and SFR coverage, abundance calculations, and
empirical fitting functions \citep[e.g.,][]{lara-lopez+10, mannucci+10,
mannucci+11, andrews+13, lara-lopez+13, delosreyes+15}.   For example,
\citet{lara-lopez+10} used 32,575 SDSS galaxies and the \citet{tremonti+04}
metallicity calibration to argue for the existence of a fundamental plane in
$M_{*}$-SFR-$Z$ space that holds out to $z = 3.5$, and an update using a
volume-selected sample from the SDSS and the Galaxy and Mass Assembly  (GAMA;
\citealp{driver+11}) surveys obtained the same result \citep{lara-lopez+13}.
In contrast, an analysis by \citet{mannucci+10}, which used the
\citet{maiolino+08} $R_{23}$ and [N~II]/H$\alpha$ abundance estimators and a
less restrictive selection sample of these same SDSS DR7 galaxies found that 
$z \sim 0.1$ galaxies fall not on a plane, but on a second order surface that 
is non-evolving out to $z = 2.5$.   Further complicating the issue is a study 
by \citet{andrews+13}, who applied a spectral stacking algorithm to a large 
number of SDSS galaxies, in order to detect the temperature-sensitive weak 
[O~III] $\lambda 4363$ auroral line and thereby directly measure 
metallicities.  The \citet{andrews+13} $M_{*}$-SFR-Z measurements are not 
consistent with the fundamental plane of \citet{lara-lopez+13} nor with the 
fundamental surface of \citet{mannucci+10}.  While some of the differences 
between these studies can be attributed to systematic uncertainties of up to 
0.7~dex between strong line abundance indicators \citep{kewley+08}, a full
explanation of these discrepancies still does not exist.

Despite these problems, a comparison between local and high-redshift
abundance estimates can still hold meaning, as long as the measurements are
performed carefully in a self-consistent manner. As shown in
Figure~\ref{fig:lineratios}, both sets of galaxies inhabit the same regions of
emission-line space, allowing the calibrations derived in
\S\ref{sec:calibration} to be applied to both datasets in the same manner.  
This should reduce systematic errors to a minimum.

We created our $z \sim 2.1$ MSZ relation following the
procedures described above.  The metallicities of our galaxies were computed by
combining the information from our three abundance sensitive line ratios,
computing each galaxy's 2-dimensional probability distribution function in
extinction--metallicity space, and then marginalizing over extinction using the
stellar reddening measurements obtained from the UV slope, the
\citet{calzetti01} prescription relating stellar and nebular attenuation, and
the \citet{cardelli+89} extinction law.  While other prescriptions for the
stellar reddening curve and the relationship between stellar and nebular
reddening are possible \citep[e.g.,][]{buat+12, scoville+15, price+14,
reddy+15}, their use makes very little difference to the final result. Our
stellar mass determinations were obtained from SED fitting (see
\S\ref{sec:SED}), and our estimates of SFR were derived using the reddening
corrected photometric measurements of the objects' UV continua and the
\citet{kennicutt+12} SFR calibration (\S\ref{sec:SFR}).   These measurements
are summarized in Table~\ref{tab:properties}.

One common way to visualize MSZ results is by plotting metallicity as a
function of a projection in SFR and stellar mass.  Unfortunately, while this
procedure reduces the dimensionality of the data and allows for easy viewing,
it ignores much of the information by averaging over the direction of
projection.   For example, when plotting our galaxies' metallicity versus
stellar mass, both our $z \sim 2.1$ and local matched samples appear to agree
very well with the \citet{dayal+13} model, and the same is true when using the
``ideal projection'' ($\mu_{0.32} = \log M_* - 0.32 \log{\rm SFR}$) found by
\citet{mannucci+10} in the local universe.  The projection washes out any
metallicity gradient that may be present.

A better way to visualize the galaxies in three-dimensional $M_*$-SFR-$Z$
space, while correctly propagating the uncertainties associated with each
abundance estimate, is to view the $M_*$-SFR plane from ``above'' and use
color to represent metallicity.  To do this we began by binning our galaxies
into hexagonal regions with radii of 0.102~dex in $\log M_*$ and $\log{\rm
SFR}$. We then took the probability contours computed from our metallicity
analysis, convolved them with the uncertainties associated with reddening and
SFR, and color-coded each hexagon according to the expectation value for
metallicity. The resulting metallicity maps for our set of $z \sim 2.1$ {\sl
HST\/} grism-selected galaxies, and our SDSS sample of stellar mass and
star formation rate matched systems are shown in
Figure~\ref{fig:averagemetallicity}.

The results of Figure~\ref{fig:averagemetallicity} can be interpreted using
some of the theoretical models that have been developed to explain a
non-evolving relationship between $M_{*}$-SFR-$Z$ \citep[e.g.,][]{dave+12,
dayal+13, lilly+13, pipino+14,forbes+14}.  Perhaps the most instructive of
these is the simple, redshift-independent analysis by \citet{dayal+13}, whose
instantaneously-recycling model takes into account star formation, inflows from
the metal-poor IGM gas, and outflows of metal-enhanced ISM gas. The basic
assumptions of this model are that star formation is proportional to gas mass,
outflows are driven by star formation via winds and supernovae, and that
inflows increase gas mass and hence trigger star formation.   There are two
main parameters of this model --- the inflow rate and the outflow rate --- and
factors of proportionality depend on halo mass, which to first order is
proportional to stellar mass \citep{dayal+09}. One way to interpret this model
is therefore through perturbation theory, since the star formation rate is
small compared to the timescale over which gas dynamics takes place.   Mergers
are not included in this analysis,  as \citet{mannucci+10} has argued that a
large number of such events would produce a much higher scatter about the
mass--metallicity relation than is observed locally.   The predictions of this
model are also shown in Figure~\ref{fig:averagemetallicity}.

>From Figure~\ref{fig:averagemetallicity}, it is clear that all three panels
confirm the existence of the well-known correlations between mass and
star formation rate, and mass and metallicity.  The local mass- and SFR-matched
sample of vigorously
star-forming SDSS galaxies also exhibits a metallicity gradient that is
approximately perpendicular to the $M_*$-SFR relation:  objects with high
stellar mass and low SFR have a higher than average metallicity.   This trend
is easily seen in the \citet{dayal+13} model, but it is not obviously visible
in our grism-selected sample.  Instead, the metallicity gradient at $z \sim
2.1$ is roughly parallel to the $M_*$-SFR relation, with higher metallicities
associated with systems with higher masses and higher star formation rates.

Another way of seeing the gradient is to subtract the metallicity of each
hexagonal bin in the local matched sample from that calculated for the $z \sim
2.1$ galaxies. This is shown in the top-left of
Figure~\ref{fig:averagemetallicitydiff}, where only those bins with at least
one galaxy in each sample have been plotted.   The noise in the map comes
principally from the {\sl HST\/} grism data, which is of lower quality than the
local SDSS measurements.  Nevertheless, there is a clear pattern to the data:
at stellar masses above $\sim 10^9 \, M_{\odot}$ and star formation rates above
$\sim 10 \, M_{\odot}$~yr$^{-1}$, the $z \sim 2.1$ galaxies tend to have higher
oxygen abundances than their local counterparts.  In contrast, in the low-mass,
low-SFR region of the diagram, most of the $z \sim 2.1$ bins have lower
abundances than their corresponding nearby objects.

Figure~\ref{fig:averagemetallicitydiff} demonstrates that the same
qualitative pattern is present when the relations of \citet{maiolino+08} are
used to determine metallicities, or if H$\beta$ luminosity is used to
measure star formation rate in our $z \sim 2.1$ systems.  
Since the set of local mass- and SFR-matched SDSS
systems described in \S\ref{sec:matches} is a better
match to the $z \sim 2.1$ grism-selected galaxies than the general ensemble of 
SDSS systems, we put more confidence in our new metallicity 
calibration than in the \citet{maiolino+08} relations, but both produce the
same trends.  Similarly, because of the generally low signal-to-noise of
our $z \sim 2.1$ H$\beta$ measurements, and because H$\beta$ SFRs are 
susceptible to shifts in metallicity \citep{zeimann+14}, we prefer to use
star formation rate measurements which are based on rest-frame UV photometry.
But this choice does not affect our results.

The top panel of Figure~\ref{fig:residuals} displays these same data, along
with their statistical error bars, as a function of mass-specific star
formation rate.  Once again, one can see the systematic behavior with mass, as
galaxies above $\sim 10^{9} \, M_{\odot}$ tend to lie in the upper part of the
figure, while those with masses below $\sim 10^{8.6} \, M_{\odot}$ mostly have
lower metallicities. However, there is considerable noise in the diagram, as
there are a limited number of galaxies in the samples, and certain sections of
the mass-SFR-metallicity phase space are poorly populated.  To help remedy this
situation,  the lower panels of Figure~\ref{fig:residuals} difference the $z
\sim 2.1$ and local galaxy results from the physically motivated
\citet{dayal+13} model that is well-calibrated locally using massive galaxies
with low star formation.  Since we use the strong-line metallicity indicators
of \S\ref{sec:calibration}, about half of the scatter in the figure can be
explained by the imperfect nature of the relations.   Nevertheless, for the
local sample, it is clear that the \citet{dayal+13} model does a reasonably
good job at predicting metallicity, apart from a few outliers.

At $z \sim 2.1$, however, the situation is different.  There is a strong
correlation between the metallicity excess over the \citet{dayal+13} model and
the mass specific star formation rate.    Galaxies which had intense
star formation $\sim 3$~Gyr after the Big Bang have a higher gas-phase
metallicity than their local counterparts.   In addition, the high-mass
galaxies of the epoch tend to have a metallicity over and above that predicted
by the models, whereas the low-mass systems have fewer metals.  Such offsets
are not seen in the local universe.

Can the differences seen in Figures~\ref{fig:averagemetallicity} through
\ref{fig:residuals} be explained by systematic offsets caused by our choice of
metallicity calibration, star formation rate indicator, or stellar mass
estimator?  As Figure~\ref{fig:averagemetallicitydiff} illustrates, the same 
patterns are present when the relations of \citet{maiolino+08} are used 
instead of our own metallicity relations.   Since our local calibration sample
is a better match to the $z\sim 2.1$ grism-selected systems, we put more
confidence in the plots of Figure~\ref{fig:averagemetallicitydiff} where we 
use the new strong-line calibrations from Table~\ref{tab:relations}, but these 
relations are not the source of the change in direction of the gradient.   
Similarly, our use of UV-based star formation rates for the $z \sim 2.1$ 
dataset does not qualitatively change our results. H$\beta$ is usually weak in 
our grism-selected galaxies, and its measurement may be affected by underlying
Balmer absorption. (\citet{brinchmann+04} have argued that the derived
SFRs for low-metallicity SDSS galaxies are still very good, even if this effect
is neglected.  Nevertheless, we have corrected our H$\beta$ emission-line 
fluxes using absorption line estimates derived from our best-fitting SED
models.)  If we then apply the same SFR conversion used for the SDSS galaxies
\citep{brinchmann+04}, we still see the same trends in the data. In order to
attribute the differences in the $M_*$-SFR-$Z$ to errors in the star formation
rate, the SFR in high-mass galaxies would need to be overestimated by no less
than $\sim 0.5$~dex, while in low-mass galaxies, SFR would need to be
underestimated by the same amount. This is exceedingly unlikely.

Analyzing the effect of errors in our stellar mass measurements is more
difficult. As noted in \S\ref{sec:matches}, the SDSS stellar masses
adopted for the local sample were based on 5-band $ugriz$ photometry, whereas
our $z \sim 2.1$ mass measurements used the entire rest-frame UV through
rest-frame IR SED\null. So systematic offsets between the two samples are
possible.  However, in order to attribute the differences in the $M_*$-SFR-$Z$
relation to this difference, a simple shift in stellar mass is not sufficient:
one would need to have the $z \sim 2.1$ galaxy masses underestimated by a 
factor $\sim 10$ for high-mass galaxies, and overestimated by a similar factor 
for low-mass systems.  Again, such errors are unlikely, though we would need a 
congruent local dataset with complete UV through near-IR photometry to 
completely exclude the possibility.

Finally, we need to consider whether the systematic trends observed in
Figure~\ref{fig:averagemetallicity} are due to changes in the physical
conditions within the epochs' H~II regions.  Because of the link between metal
abundance and stellar opacity, there is a well-known correlation between the 
ionization parameter and metallicity \citep[e.g.][and references 
therein]{kewley+13}.  This, in part, defines the behavior of the strong-line 
metallicity indicators used in this paper.  However, if this correlation 
changes between $z \sim 2.1$ and today, then our metallicity determinations 
could be in error. Again, this shift would need to cause us to overestimate 
the gas-phase abundances for our $z \sim 2.1$ high-luminosity, high-metallicity
systems, while underestimating $Z$ in low-metallicity  galaxies.  We attempted 
to minimize this effect by creating a sample of local galaxies with
physical conditions as close to those at $z \sim 2.1$ as possible.
Nevertheless, we cannot fully rule out this possibility.

Despite our best effort to reduce systematics, we still see correlations
between the metallicity offsets in stellar mass, SFR, and sSFR\null.  On
average over a large enough region of phase space, these offsets cancel out,
but as it stands, the $M_*$-SFR-$Z$ surface defined by our $z \sim 2.1$ {\sl
HST\/} grism-selected galaxies is different from that of our local mass- and
SFR-matched SDSS sample, and different from the physically-motivated model of
\citet{dayal+13}.

%%%%%%%%%%%%%%%%%%%%%%%%%%%%%%%%%%%%%%%%%%%%%%%%%%%%%%%%%%%%%%%%%%%%%
\section{Summary}
\label{sec:summary}

In this study we examined the distribution of intensely star-forming $z \sim
2.1$ galaxies in stellar mass--SFR--metallicity space and compared our results 
to a sample of local galaxies.  This study differs from previous investigations
of this Fundamental Metallicity Relation in that it contains significant 
numbers of galaxies with low stellar mass ($7.5 \lesssim \log(M/M_{\odot}) 
\lesssim 10$), and high star formation rates $-0.5 \lesssim \log {\rm SFR} 
\lesssim 2.5$ (in $M_{\odot}$~yr$^{-1}$).  In total, we have 256 measurements 
of stellar mass, metallicity, and star formation in the $z \sim 2.1$ universe.
To summarize our findings:

\begin{enumerate}

    \item We calibrate the strong-line metallicity indicators using a local
        sample of emission line galaxies
        with high H$\beta$ luminosities
        ($L_{{\rm H}\beta} \geq 3 \times 10^{40}$~ergs~s$^{-1}$) and matched in
        stellar mass and star formation rate to our $z \sim 2.1$ sample.
        These galaxies obey
        the \citet{maiolino+08} metallicity calibrations except at low
        metallicity, where our relations produce abundances that are lower by
        $\sim 0.15$~dex.  This makes a difference for galaxies with masses
        below $\sim 10^{8.5} \, M_\odot$ (\S\ref{sec:calibration}).

    \item By combining several strong-line ratio metallicity indicators
        ($R_{23}$, [O~III]/[O~II], and [Ne~III]/[O~II]) into a single
        meta-indicator, we break the well-known degeneracy in the $R_{23}$
        relation; this is important, since many of our galaxies lie close to
        the peak of the $R_{23}$ curve.  The combination of indicators also
        significantly reduces the statistical error in our measurements.

    \item In a metallicity map in $M_*$-SFR space
        (Figure~\ref{fig:averagemetallicity}), the metallicity gradient for the
        $z \sim 2.1$ {\sl HST-}selected galaxies is along the $M_*$-SFR
        relation, whereas for the local galaxies, the gradient is roughly
        perpendicular to this correlation.

    \item At masses above $\sim 10^9 \, M_{\odot}$ and star formation
        rates above $\sim 10 \, M_{\odot}$~yr$^{-1}$, star-forming
        galaxies at $z \sim 2.1$ have higher metallicities than similar
	objects in the local universe.  
        At lower masses and star formation rates,
        the $z \sim 2.1$ systems tend to have lower metallicities than their
        local counterparts (Figure~\ref{fig:averagemetallicitydiff}).

    \item At $z \sim 2.1$, intensely star-forming galaxies have higher
        metallicities than predicted by the chemical evolution model of 
        \citet{dayal+13}.  Similar local galaxies only deviate from 
        this model at very low specific star formation rates
        (Figure~\ref{fig:residuals}).
\end{enumerate}

\acknowledgments
This work was supported via NSF through grant AST 09-26641\null.  The
Institute for Gravitation and the Cosmos is supported by the Eberly College of
Science and the Office of the Senior Vice President for Research at the
Pennsylvania State University. We thank the referee whose valuable comments
greatly improved this paper. We also thank Viviana Acquaviva for the use of her
SED fitting code {\tt GalMC}.

{\it Facilities:} \facility{HST (WFC3)}

\clearpage

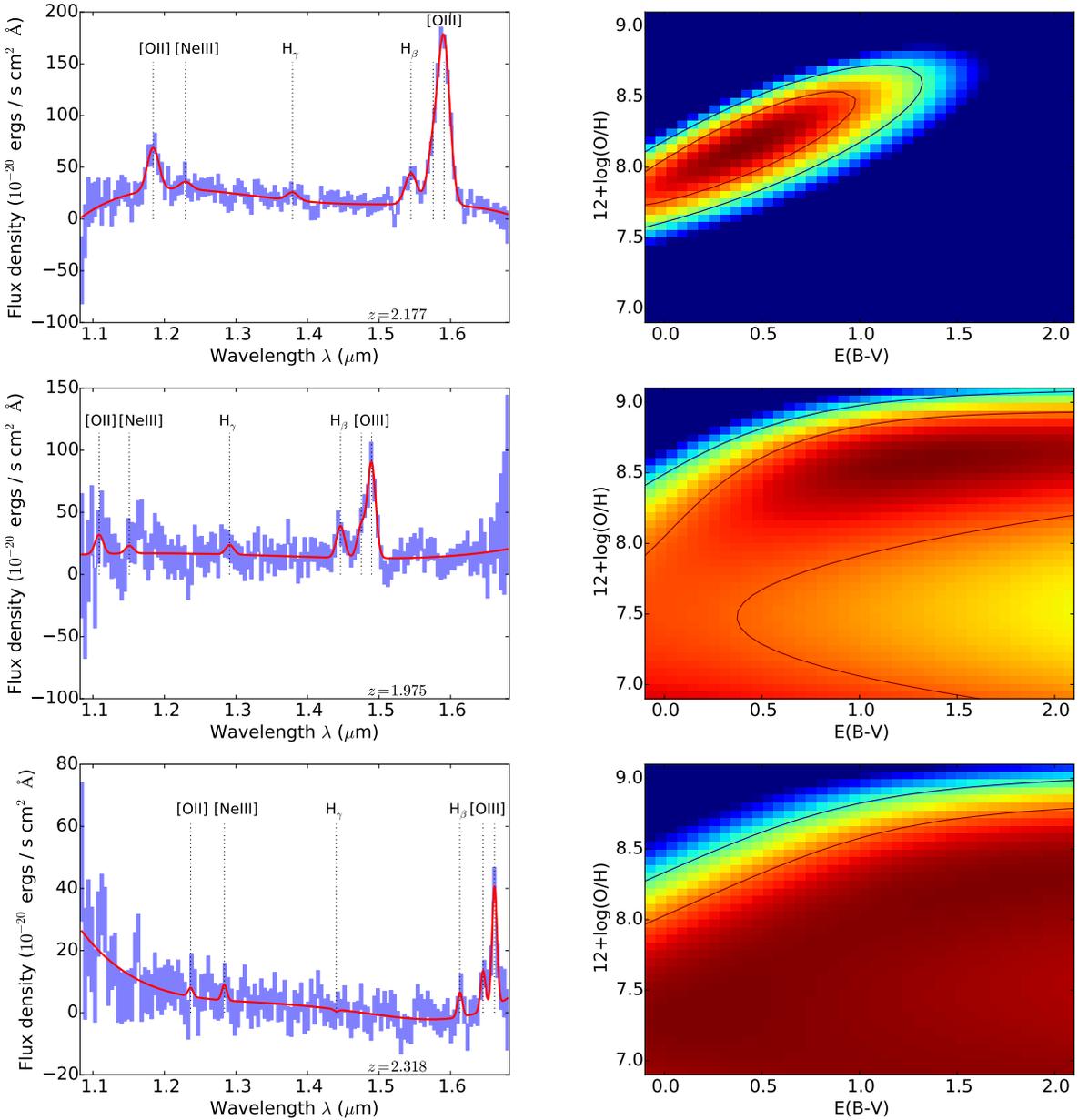
\begin{figure}[t]
\newcommand{\specfits}[2]{%
  \node at (0,#2) {\includegraphics[width=\mywidth,page=#1]{spectra.pdf}};
  \node at (\mywidth,#2) {\includegraphics[width=\mywidth,page=#1]{zvsebv.pdf}};
}
\makebox[\textwidth]{
\begin{tikzpicture}
  \def\offset{-5.5}
  \def\mywidth{0.5\textwidth}
  \specfits{1}{0}
  \specfits{2}{\offset}
  \specfits{3}{\offset+\offset}
\end{tikzpicture}
}
\caption{Three galaxies that exemplify the range of fitting solutions for our
sample of $z \sim 2.1$ grism-selected galaxies. The spectra on the left are 
shown with 1-$\sigma$ error bars, with the best-fitting model illustrated as a
solid red line. The metallicity vs.\ $E(B-V)$ contours are shown on the right,
marginalized over all nuisance parameters.  In many cases, the metallicity
is moderately well-determined, even if the reddening is not; in other
cases, an estimate of extinction would greatly improve the abundance
measurement.  While we cannot measure nebular extinction directly, we can 
constrain $E(B-V)_{\rm gas}$ via measurements of stellar reddening and 
the \citet{calzetti01} attenuation law.}
\label{fig:spectra}
\end{figure}

\begin{figure}[t]
    \centering
    \includegraphics{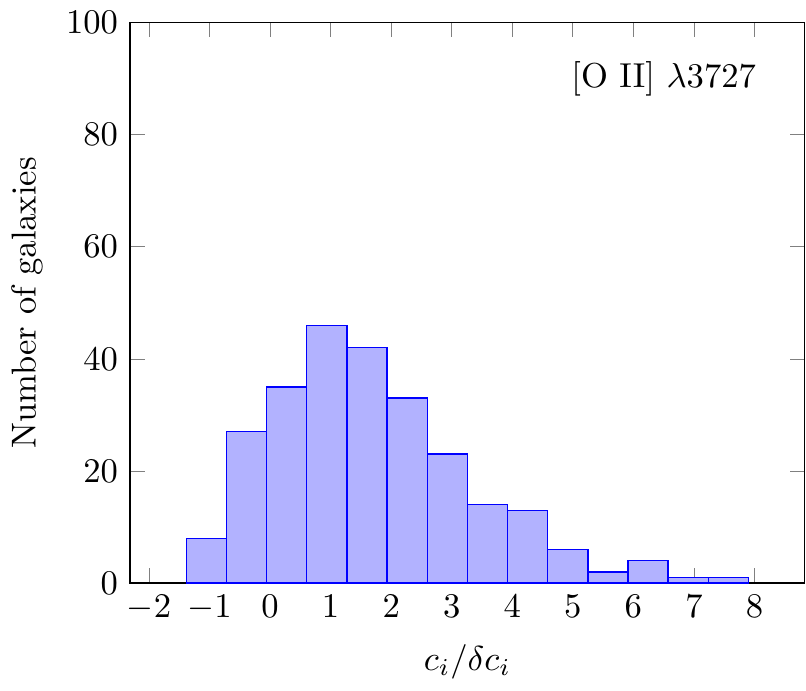}
    \includegraphics{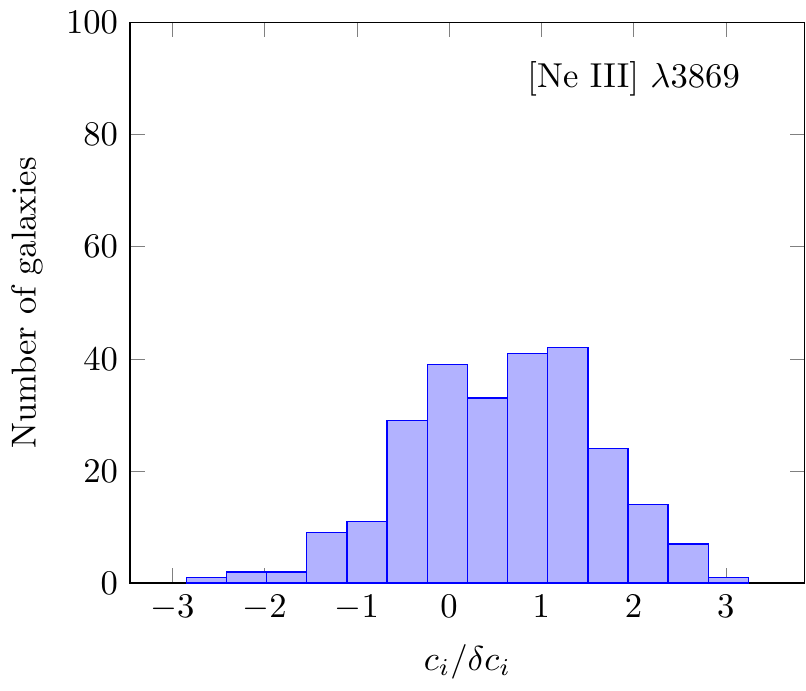}
    \includegraphics{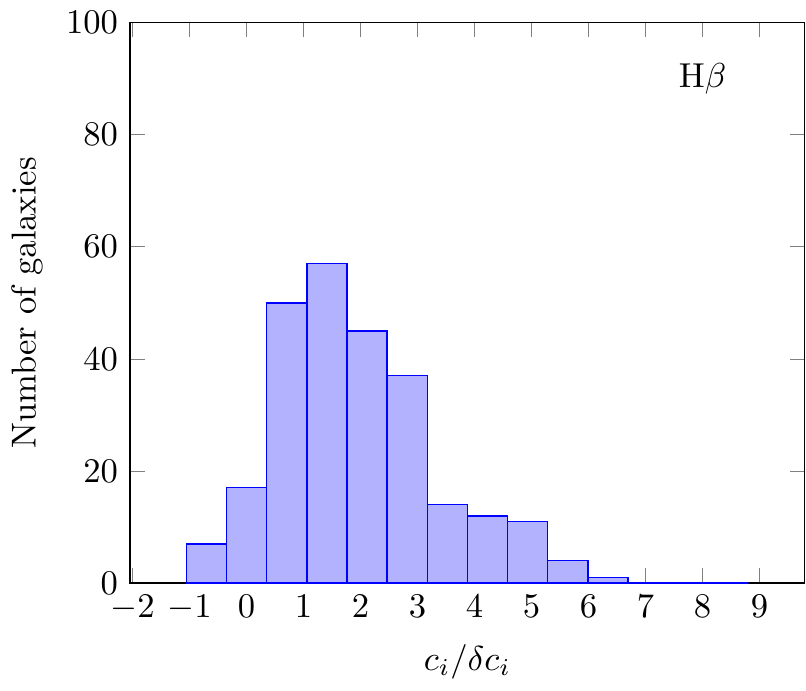}
    \includegraphics{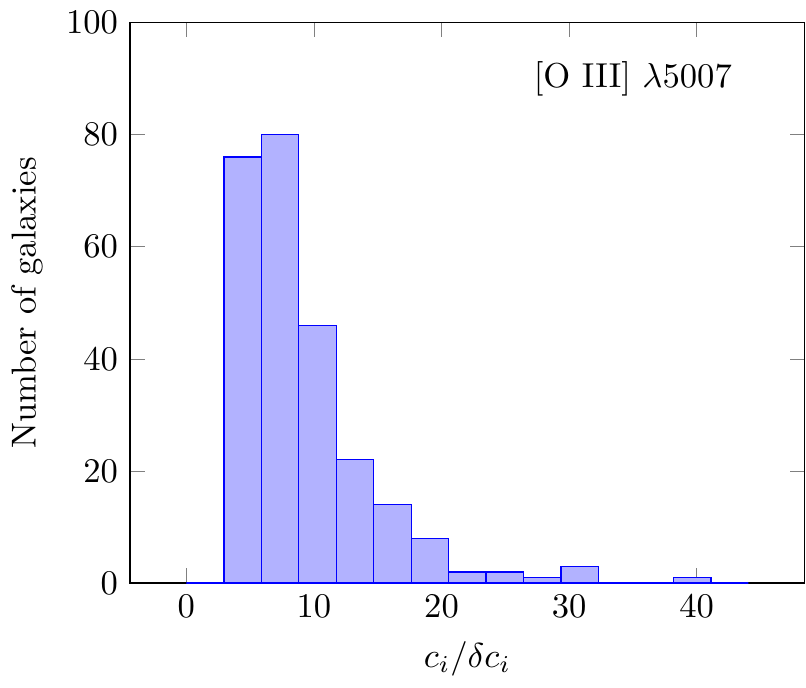}
    \caption{
        \changed{
        Histograms for the ratio of measured line flux $c_i$ to uncertainty
        $\delta c_i$ for [O~II]~$\lambda3727$, [Ne~III]~$\lambda3869$,
        H$\beta$, and [O~III]~$\lambda5007$ for our $z\sim2.1$ galaxies. Since
        these systems were selected via the blended [O~III] doublet, the
        minimum $c_i/\delta c_i$ for [O~III] is three. Since the
        signal-to-noise is quite low for some of our galaxies, stochasticity
        demands that some of the measured fluxes $c_i$ are negative. Thus,
        measuring metallicity can be quite challenging. Nevertheless, because
        the redshift is fixed by [O~III], line fluxes can be measured to much
        lower signal-to-noise, although with large statistical uncertainties.
        Our Bayesian approach to metallicity determinations (see
        \S\ref{sec:metallicity}) accounts for this uncertainty, and we
        propagate the full p.d.f. of each measurement into our final analysis.
        }
    }
    \label{fig:snr}
\end{figure}

\begin{figure}[t]
\ifpdf
    \centering
    \includegraphics{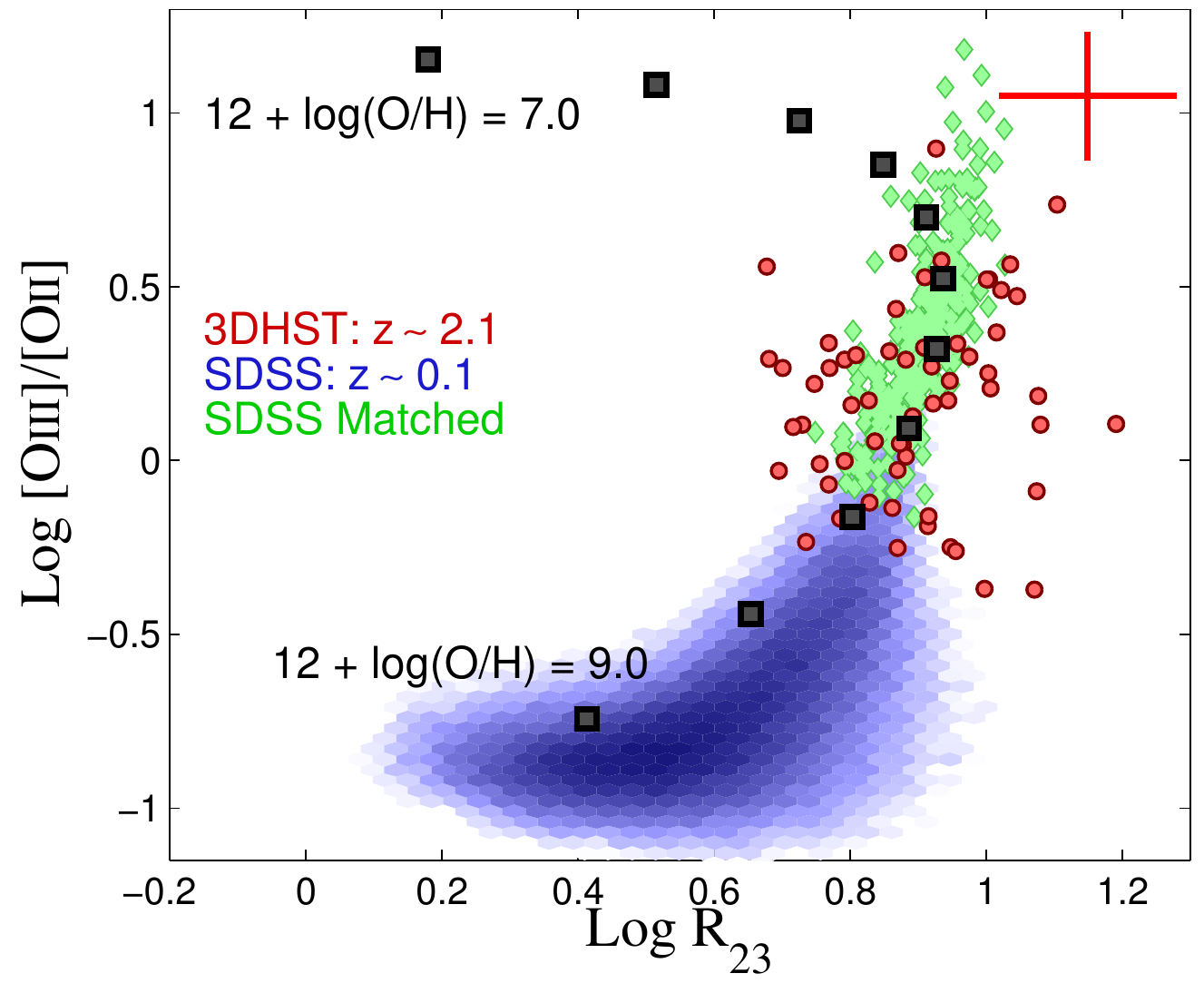}
\else
    \epsscale{0.8}
    \plotone{r23_o3hb.eps}
\fi
\caption{The relationship between two of the three line ratios used to
empirically determine gas-phase metallicity for our sample of $z \sim 2.1$
galaxies.  The nearly 200,000 star-forming galaxies from the SDSS DR7 catalog
are shown via the blue density plot, while our sample of $z \sim 2.1$
grism-selected galaxies is shown in red.  (Only those galaxies where all four
of our key emission lines have detections with S/N $> 4$ have been plotted).
The red error bar in the upper right corner is the median error of this sample.
A local calibration sample, i.e., systems with H$\beta$ luminosities
larger than $3 \times 10^{40}$~ergs~s$^{-1}$ and matched in stellar mass and
SFR to our $z\sim 2.1$ systems are shown as green diamonds.   For
comparison, the black squares show the empirical relation of
\citet{maiolino+08}, which extends from $12+\log({\rm O/H}) = 9.0$ to 7.0 in
0.2 dex intervals.  Within the measurement errors, our galaxies occupy the same
emission-line phase space as the local calibration sample, while the
\citet{maiolino+08} calibrations deviate significantly from this locus at low
metallicity.}
\label{fig:lineratios}
\end{figure}

\begin{figure}[t]
%\epsscale{.7}
\plotone{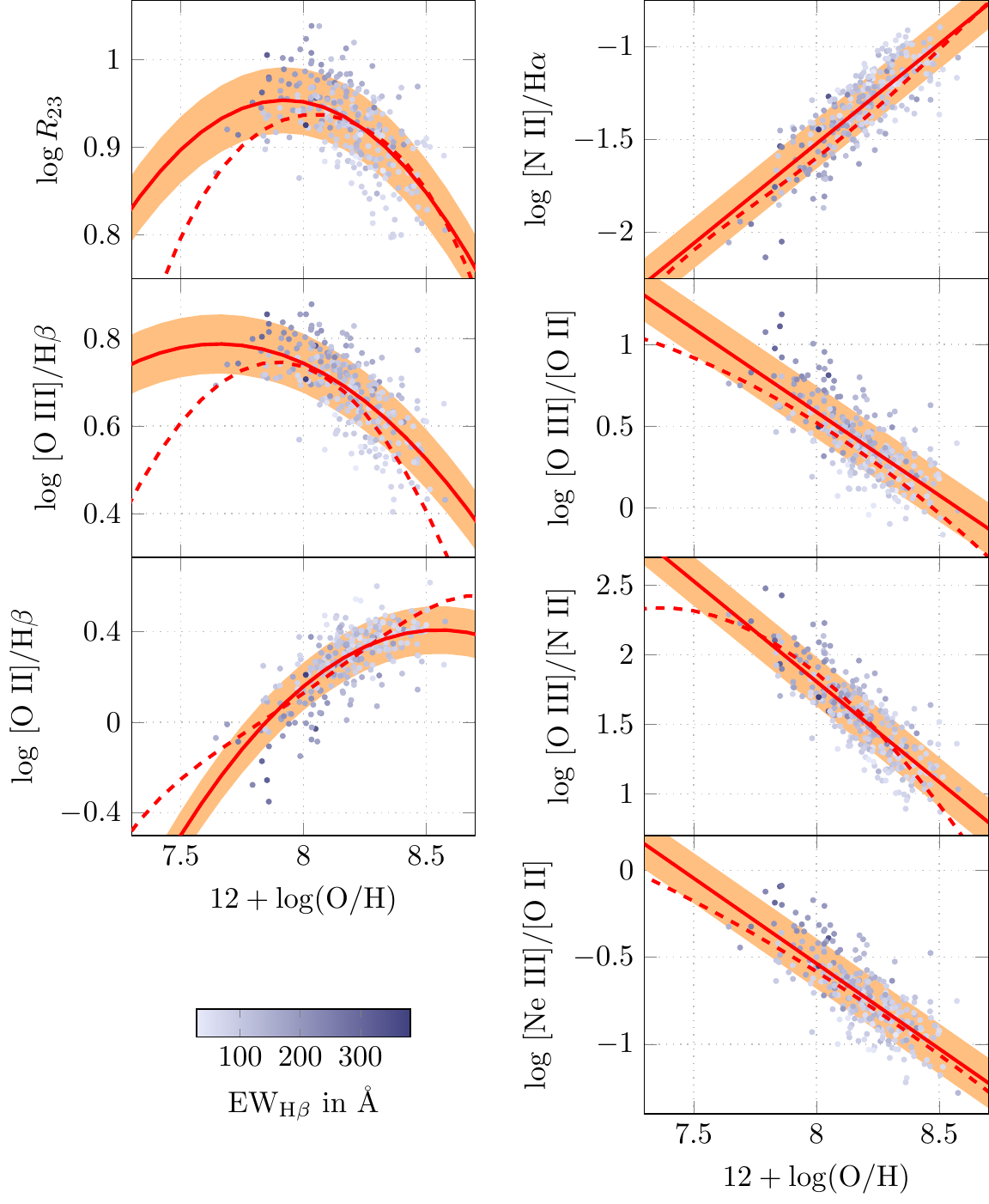}
\caption{The gas-phase oxygen abundances versus strong emission-line ratios for
our local calibration sample of high luminosity ($L_{H\beta} > 3 \times
10^{40}$~ergs~s$^{-1}$), mass- and SFR-matched galaxies.
The objects are color-coded by their rest-frame equivalent width. The
solid red lines show the polynomial fits for our calibration sample, with the
scatter indicated by the orange-shaded area (Table~\ref{tab:relations}). The
dashed red lines are the \citet{maiolino+08} relations.}
\label{fig:newmaiolino}
\end{figure}

\begin{figure}[t]
\epsscale{0.9}
\plotone{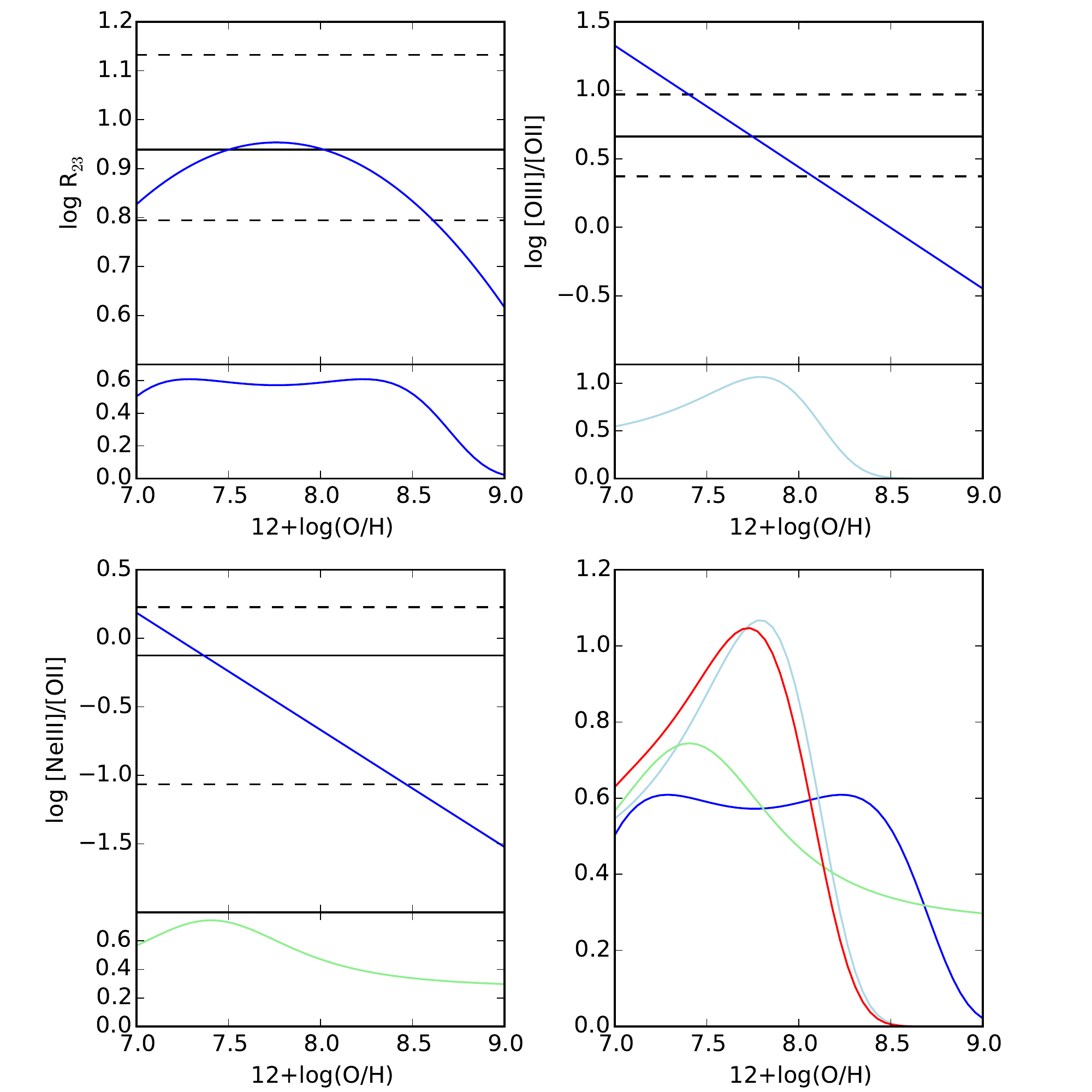}
\caption{A demonstration of how the different indicators contribute to the
metallicity estimate for one of the galaxies in our sample.  The lower curves
in the upper left ($R_{23}$), upper right ([O~III]/[O~II]), and lower left
([Ne~III]/[O~II]) panels show the likelihood distributions for each indicator,
normalized in the range $Z=7-9$.  These were derived by predicting the relative
line fluxes using the polynomial relationships between line ratio and
metallicity (upper solid blue curve), and then comparing these ratios to the 
measured best-fit line fluxes and their associated uncertainties.
The solid black line shows the best fit ratio,
with the 16th and 84th-percentiles illustrated as dashed lines.  The bottom
right panel repeats these probability distributions ($R_{23}$ in blue,
[O~III]/[O~II] in teal, and [Ne~III]/[O~II] in green) and shows
the probability distribution function for the combined three-ratio 
meta-indicator in red.}
\label{fig:metal_demo}
\end{figure}

\begin{figure}[t]
    \centering
    \input zcompare_matched.tex
    \caption{
        The metallicities obtained using our meta-indicator of three line
        ratios, $R_{23}$, [O~III]/[O~II], and [Ne~III]/[O~II] (see
        Table~\ref{tab:relations}) compared to physics-based abundances 
	determined using the electron temperature sensitive [O~III]
        $\lambda\lambda 4959,5007$/[O~III] $\lambda 4363$ line ratio. The
        dashed red line depicts where the abundances match.   On average,
        strong-line metallicity estimates are greater than the direct
        abundances by $0.028 \pm 0.154$~dex;  this agreement is better than
        that for the \citet{maiolino+08} relations ($-0.041 \pm 0.148$~dex),
        though the scatter is somewhat larger.  Our meta-indicator method is
        reliable and biased by no more than $\sim 0.03$~dex.
    }
    \label{fig:test_stats}
\end{figure}

\begin{figure}[t]
\centering
\includegraphics[width=0.45\textwidth]{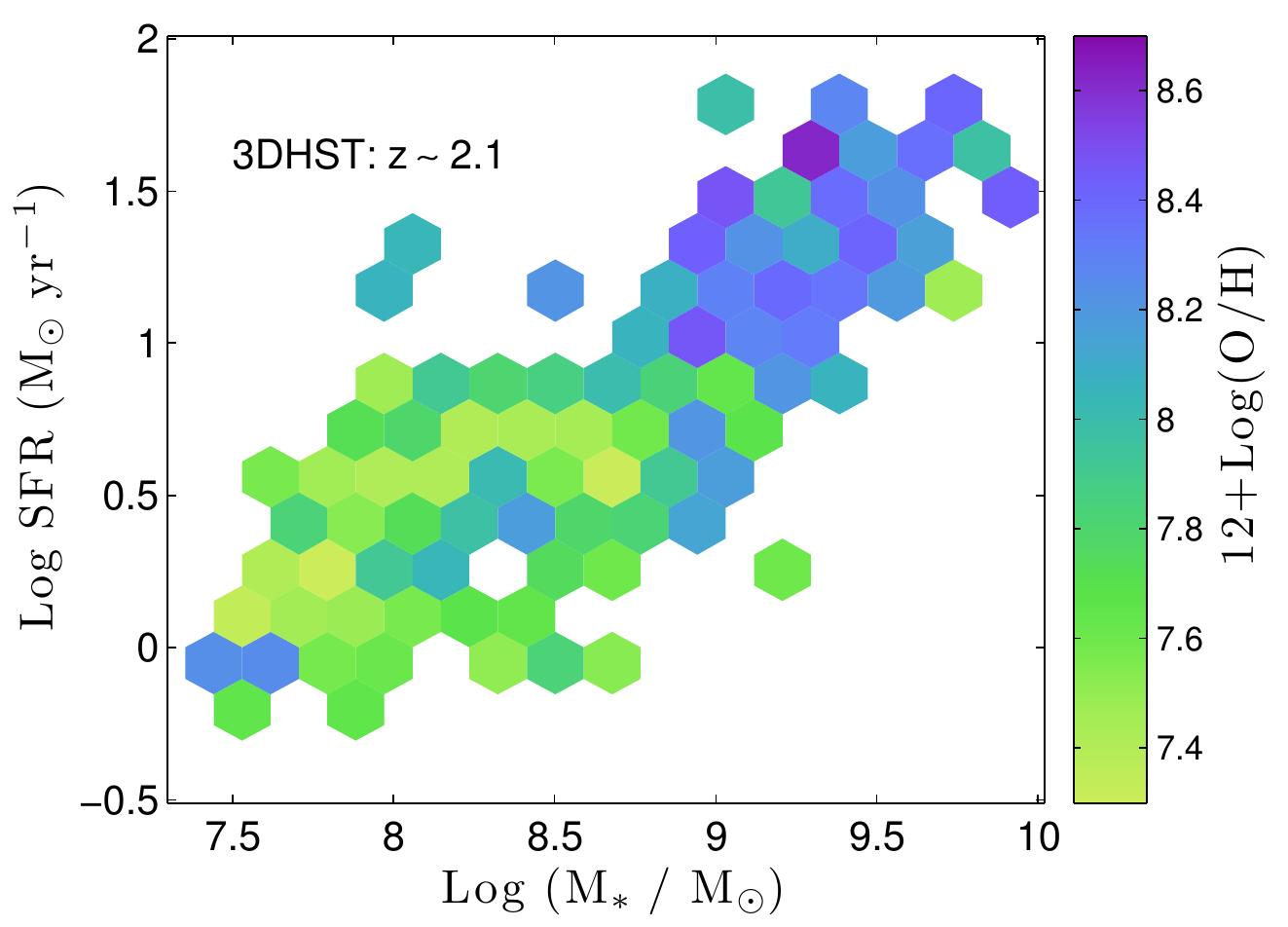}
\includegraphics[width=0.45\textwidth]{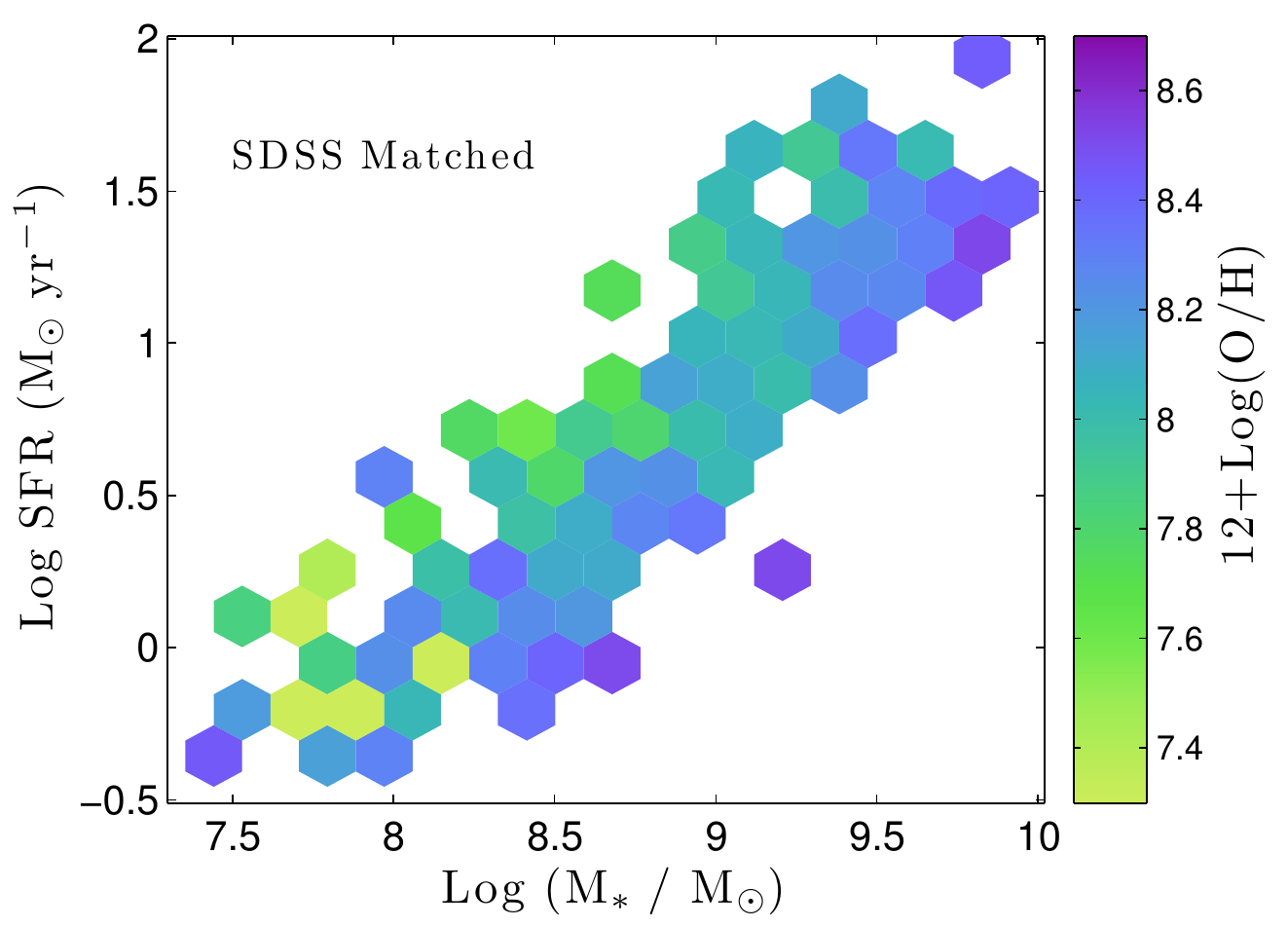}
\newline
\includegraphics[width=0.45\textwidth]{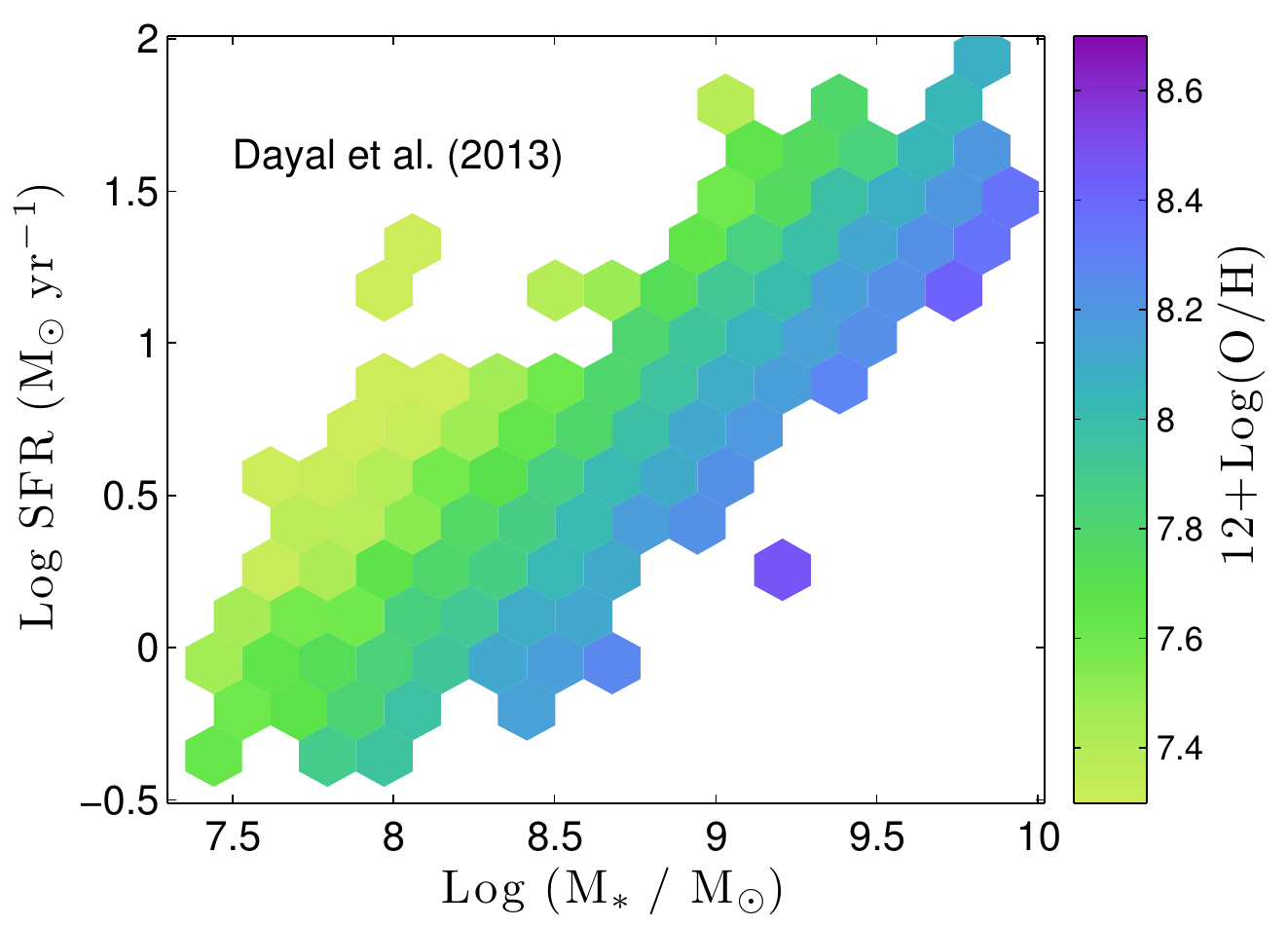}
\caption{Three panels showing the relationship between stellar mass, 
star formation rate, and metallicity for $z \sim 2.1$ galaxies (upper left), a
local sample of SDSS galaxies matched in stellar mass and star formation rate
(upper right), and the non-evolving \citet{dayal+13} chemical evolution model.
The metallicities of the two samples of galaxies have been determined in the 
same way using our new polynomial relations (see \S\ref{sec:calibration}).  
The data have been binned into 0.102~dex hexagons, and only those bins with at 
least one galaxy have been plotted.   In the local sample, there is a
metallicity gradient that runs roughly perpendicular to the $M_{*}$-SFR 
correlation that follows the predictions of the \citet{dayal+13} model. This 
gradient is in a different direction at $z \sim 2.1$.} 
\label{fig:averagemetallicity}
\end{figure}

\begin{figure}[t]
\newcommand{\plotpdf}[2][]{%
    \begin{minipage}{0.5\textwidth}
        \vspace{2mm}
        \hspace{0.3\textwidth} #1
        \newline
        \includegraphics[width=\textwidth]{#2}%
    \end{minipage}
}
\makebox[\textwidth]{
    \begin{minipage}{\textwidth}
        \begin{multicols}{2}
            \begin{itemize}
                \item Using new metallicity relations:
            \end{itemize}
            \centering
            \vspace{-3mm}
            \plotpdf[UV SFR:]{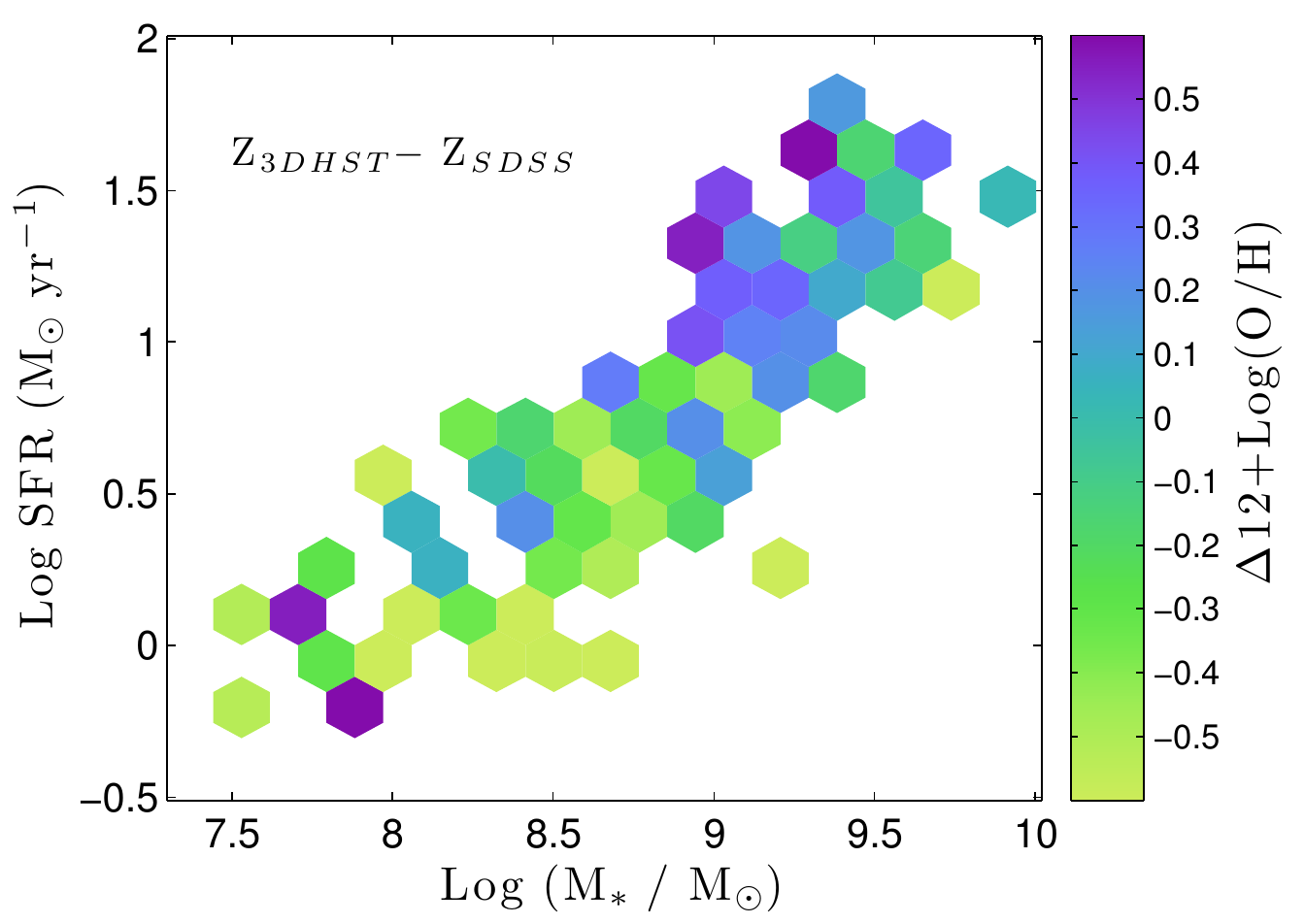}
            \plotpdf[H$\beta$ SFR:]{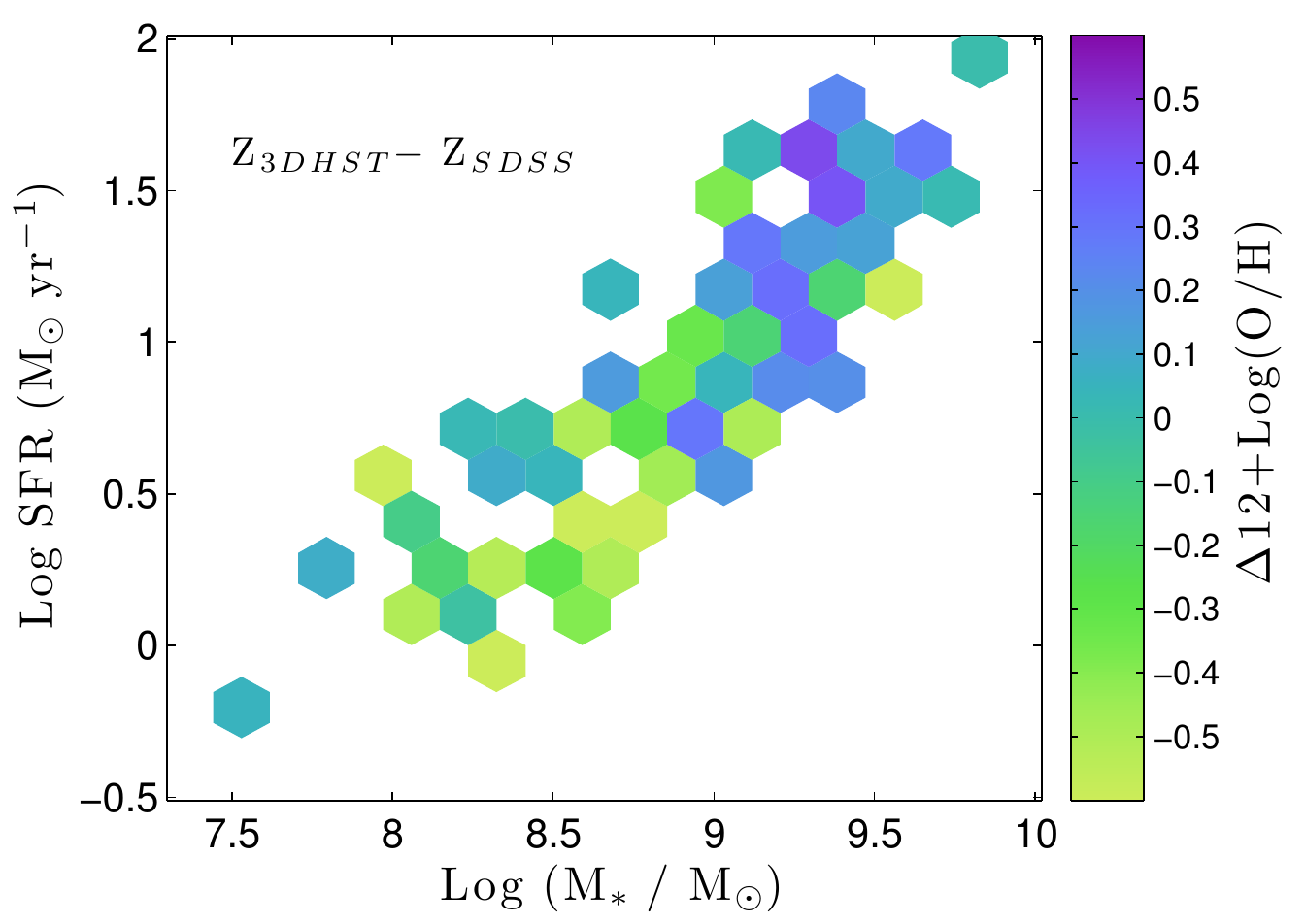}

            \begin{itemize}
                \item Using \citet{maiolino+08} relations:
            \end{itemize}
            \centering
            \vspace{-3mm}
            \plotpdf[UV SFR:]{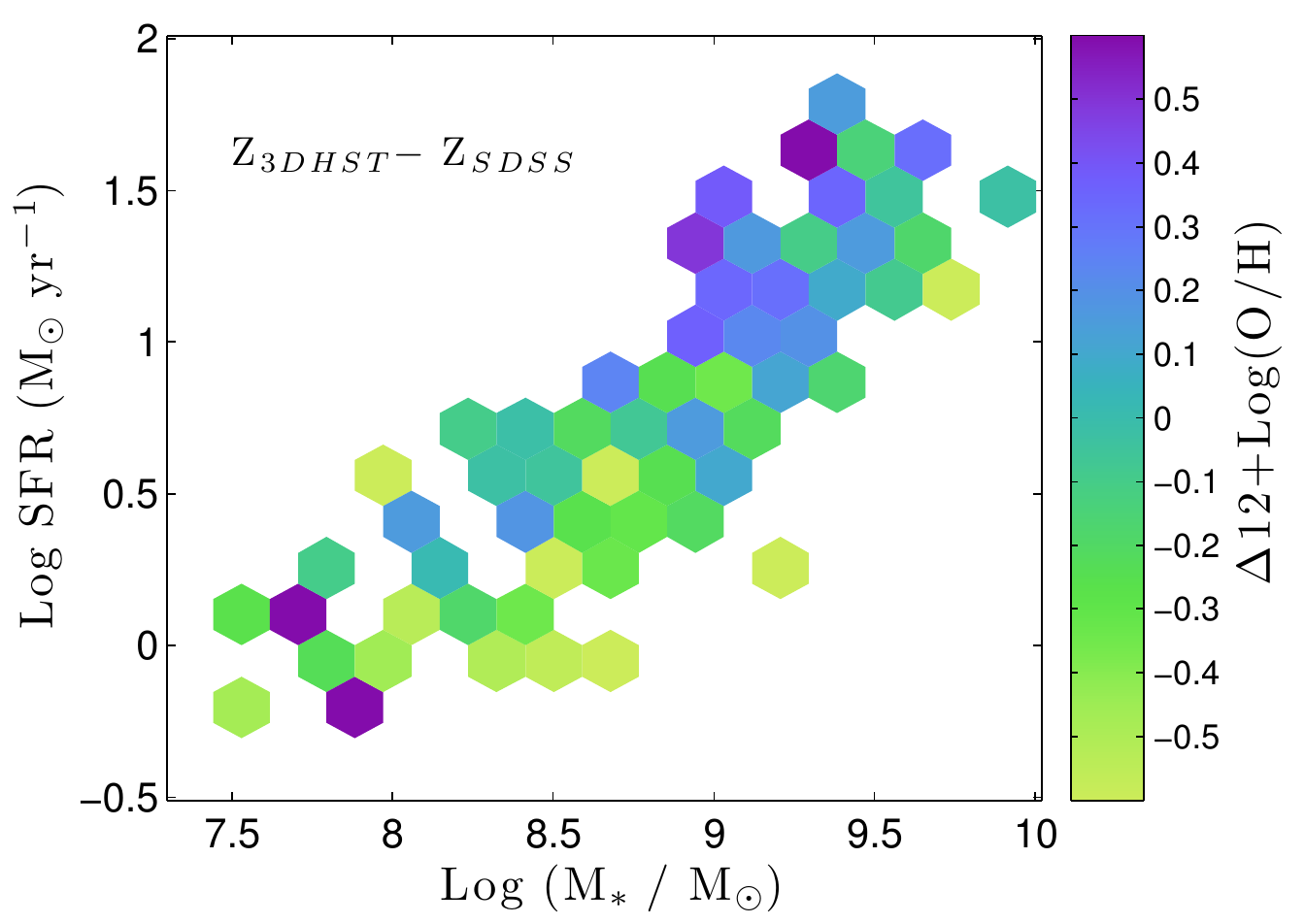}
            \plotpdf[H$\beta$ SFR:]{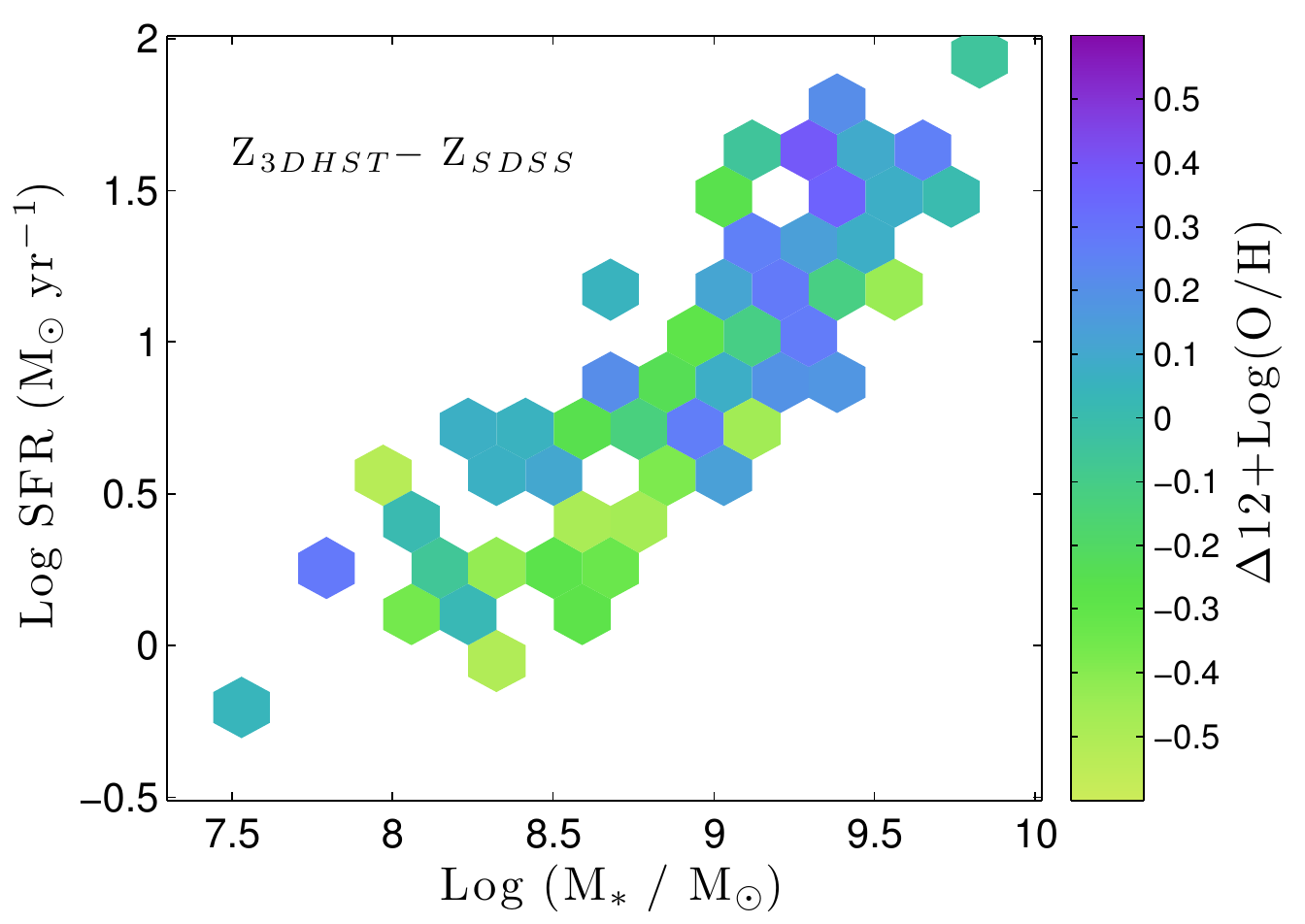}
        \end{multicols}
    \end{minipage}
}
\caption{The difference between the metallicity measurements of $z \sim 2.1$
grism-selected galaxies and similar measurements for our local sample of high 
H$\beta$ line-luminosity galaxies matched in stellar mass and star formation 
rate.  The data have been binned into 0.102~dex hexagonal 
pixel.  The left column shows abundance estimates determined via the 
metallicity calibration derived from our local sample of galaxies, with the
top panel showing the distribution using rest-frame UV photometry as our
SFR indicator and the bottom plot using H$\beta$ to estimate the star formation
rate.   The right column repeats these plots using 
the \citet{maiolino+08} relations to estimate metallicity.  All four
plots illustrate the same result:  at stellar masses above
$\sim 10^9 \, M_{\odot}$ and star formation rates above 
$\sim 10 \, M_{\odot}$~yr$^{-1}$ all but three of the bins have a higher 
metallicity at $z \sim 2.1$ than seen locally.  At lower
masses and SFRs, it is the nearby galaxies that are more metal-rich.
}
\label{fig:averagemetallicitydiff}
\end{figure}

\begin{figure}[t]
\centering
\hspace{2cm}  % LaTeX, why is this needed?
\includegraphics[width=0.45\textwidth]{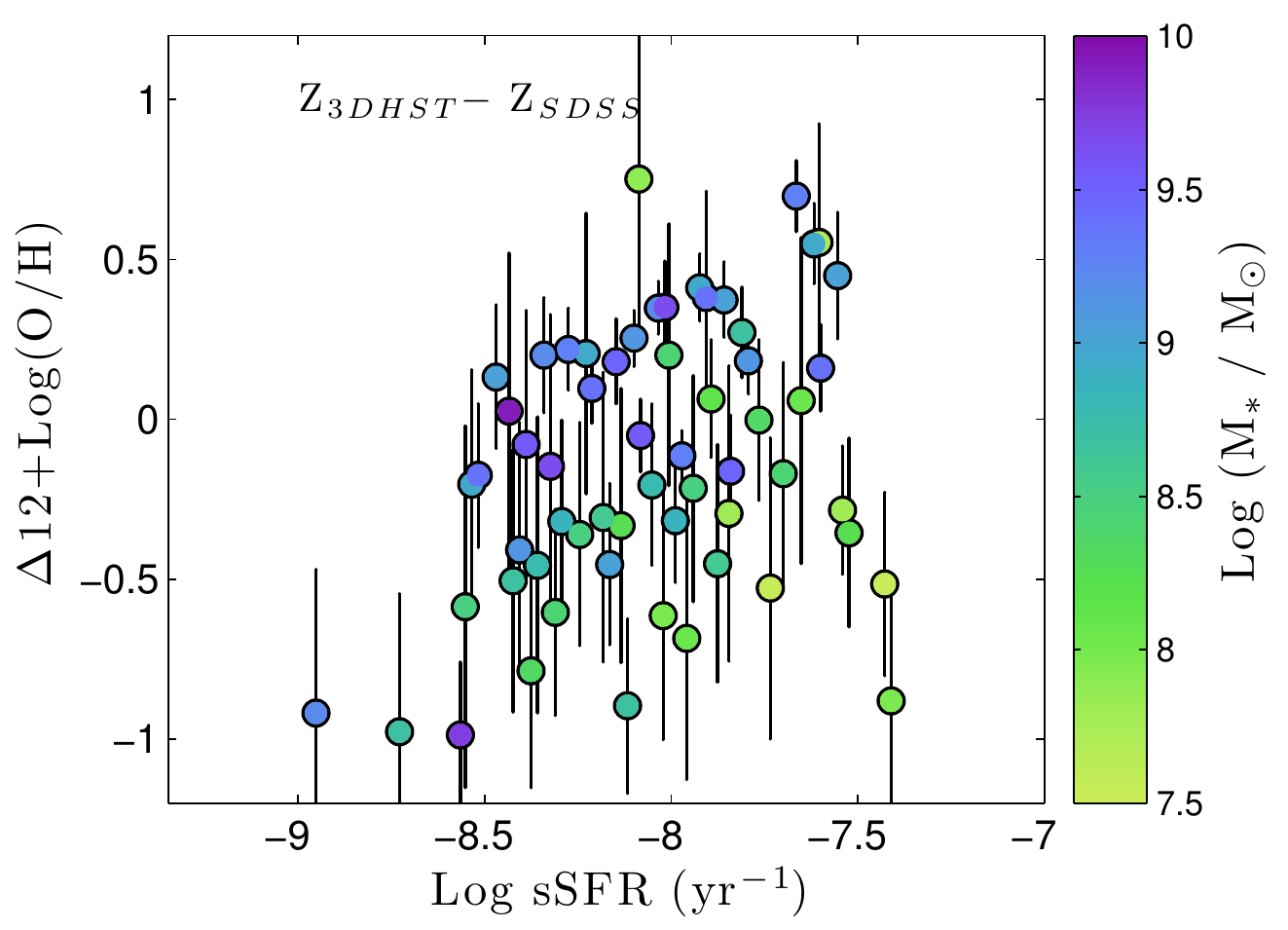}
\newline
\includegraphics[width=0.45\textwidth]{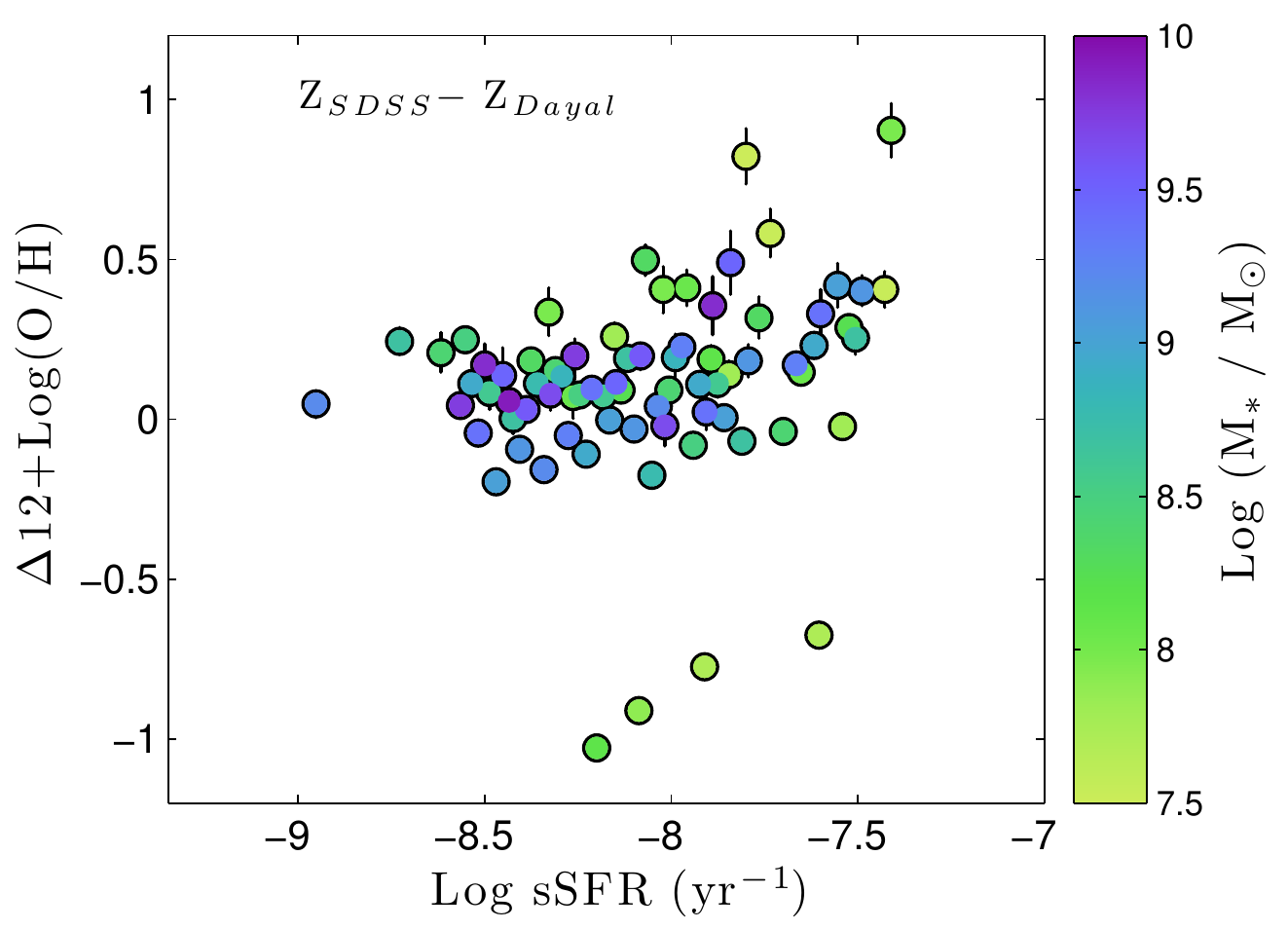}
\includegraphics[width=0.45\textwidth]{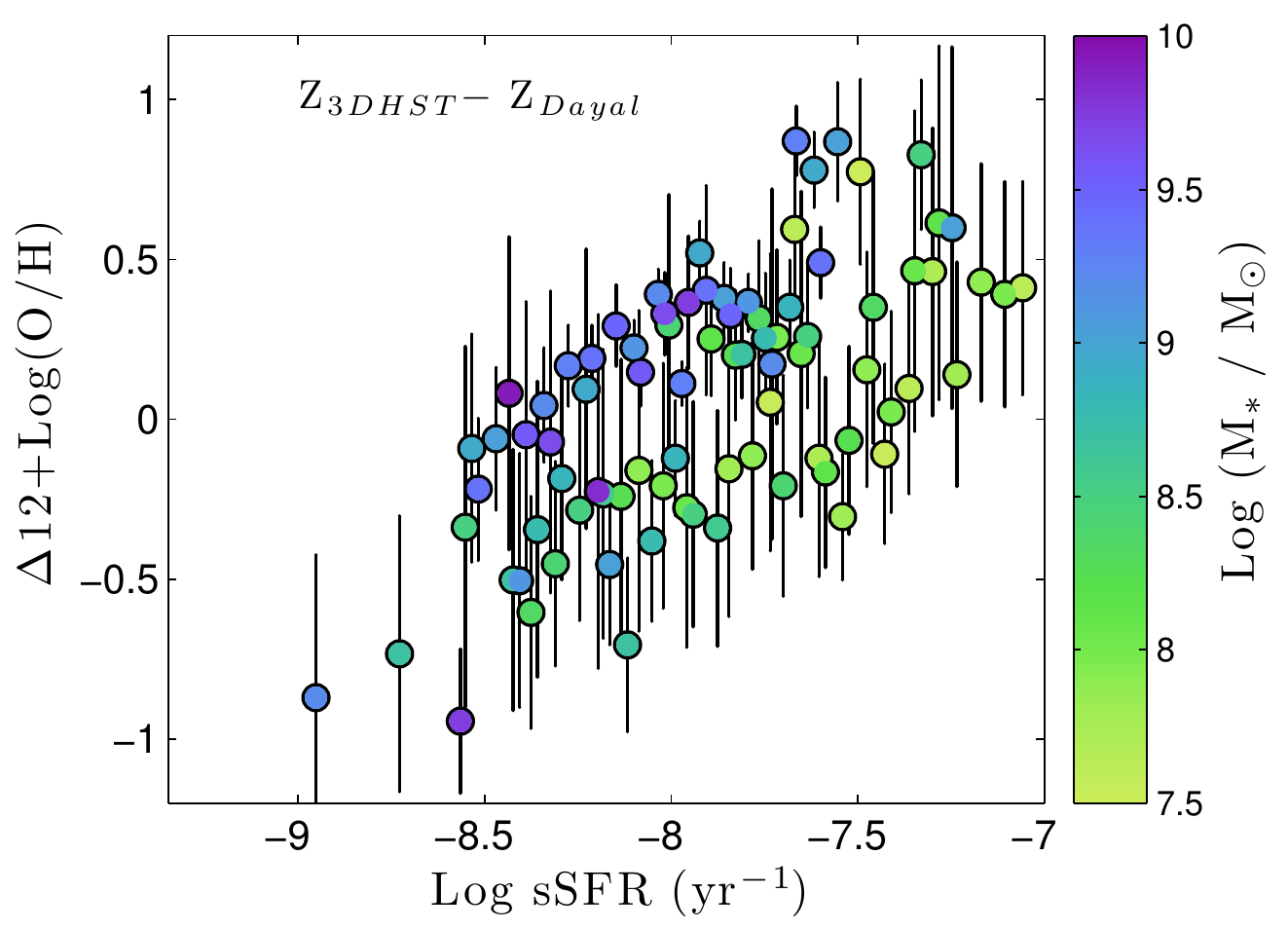}
\caption{Residual metallicities between the $z \sim 2.1$ 3D-HST sample, our 
local SDSS comparison sample of high H$\beta$ line-luminosity galaxies
matched in stellar mass and SFR, and the 
\citet{dayal+13} model, as a function of mass-specific star formation rate
(sSFR\null).  Each data point represents one of the hexagons in
Figure~\ref{fig:averagemetallicity}, colored by stellar mass. The error bars
illustrate the statistical uncertainty of each bin. \emph{Top:} Difference
between the two samples. Only bins that have at least one galaxy in each of the
two samples are shown. High-mass galaxies in the 3D-HST sample tend to have
higher metallicities than the local SDSS matches. \emph{Lower left:}  The
matched sample local counterparts compared to the \citet{dayal+13} chemical 
evolution model.  About half the scatter is explained by the dispersion about 
the metallicity calibrations (see \S\ref{sec:calibration}), and some outliers 
are present. \emph{Lower right:}  Metallicity residuals between the 3D-HST 
sample and the \citet{dayal+13} model.  As in the previous plot, the 3D-HST 
galaxies tend to have lower metallicities
at lower star formation intensities.  However, at larger sSFRs, the 3D-HST
galaxies have higher metallicities than the model.  This is not seen locally,
and is especially true for the higher-mass galaxies in the sample.}
\label{fig:residuals}
\end{figure}

\clearpage

\appendix
%%%%%%%%%%%%%%%%%%%%%%%%%%%%%%%%%%%%%%%%%%%%%%%%%%%%%%%%%%%%%%%%%%%%%%%%%%%%%%
\section{Metallicity via Maximum Likelihood}
\label{appendix:a}
Here we detail the maximum-likelihood method for obtaining line 
fluxes from a grism spectrum.  We then detail our maximum-likelihood method for
obtaining metallicity using strong-line indicators with intrinsic scatter.

\subsection{Line Fluxes}
We model each 1-D extracted spectrum $f(\lambda)$ as a sum of $k = 6$ Gaussian
lines superimposed on a polynomial continuum of order $n = 4$.  In other
words, the flux $f$ as a function of wavelength $\lambda$ is
\begin{equation}
    f(\lambda) = \sum_{i=0}^n c_i \lambda^i + \sum_{j=n+1}^{n+k} \frac{c_j}{\sqrt{2 \pi}
\sigma} \exp \left( - \frac{(\lambda - \lambda_{e,j})^2}{2 \sigma^2} \right) +
w(\lambda)
\label{eq:a_model}
\end{equation}
where $c_i$ are the coefficients to the polynomial associated with the
continuum, $c_j$ are the fluxes of the emission lines, $\sigma$ is the width of
the Gaussians, $\lambda_{e,j}$ are the redshifted central wavelengths of the
emission lines, and $w(\lambda)$ represents the random noise of the detection.
The flux falling within each pixel of size $\Delta x$ is 
\begin{equation}
F(x) = \int_{x - \Delta x / 2}^{x + \Delta x/2} f(\lambda) \, d\lambda
\label{eq:a_pixelation}
\end{equation}
where $x$ is the central wavelength of a pixel.
The resulting model can then be written as
\begin{equation}
F(x) = \sum_{i=0}^{n+k} c_i g_i(x) + W(x) 
\label{eq:a_newmod}
\end{equation}
where $W(x)$ is the noise and the resulting $g_i(x)$ depend only on 
the redshift of the galaxy, the line widths, and the size of a pixel.

The likelihood that the model $F(\vec{x})$ fits a measured spectrum $\vec y$ is 
\begin{equation}
L(c_i|\vec{y}) \propto \exp \left( - \frac{1}{2} \left( F(\vec x) - \vec y
\right)^T \Sigma_w^{-1} \left( F(\vec x) - \vec y \right) \right) 
\label{eq:a_likelihood}
\end{equation}
where $\Sigma_w$ is the matrix containing the covariance of
each point.  If the points are independent, this matrix is diagonal with
elements $1/e_i^2$, where $e_i$ are the Gaussian errors of each data point. 
Taking the natural log of the likelihood function we get
\begin{equation}
\begin{split}
\ln L(c_i,z,\sigma|\vec y) &= -\frac{1}{2} \left( \sum_i c_i g_i(\vec x) + W(\vec x) - 
\vec y \right)^T \Sigma_w^{-1} \left( \sum_j c_j g_j(\vec x) + W(\vec x) - \vec y 
\right) \\
&= -\frac{1}{2} \sum_{i,j} c_i c_j g_i(\vec x)^T \Sigma_w^{-1} g_j(\vec x) -
\sum_i c_i g_i(\vec x)^T \Sigma_w^{-1} \left( W(\vec x) - \vec y \right)  \\
&\hskip60pt
-\frac{1}{2} \left( W(\vec x) - \vec y \right)^T \Sigma_w^{-1} \left( 
W(\vec x) - \vec y \right) 
\end{split}
\label{eq:a_loglikelihood}
\end{equation}
By introducing
\begin{equation}
    A_{ij} = g_i(\vec x)^T \Sigma_w^{-1} g_j(\vec x) 
\label{eq:a_a}
\end{equation}
\begin{equation}
    B_i = -g_i(\vec x)^T \Sigma_w^{-1} \left( W(\vec x) - \vec y \right) \approx
    g_i(\vec x)^T \Sigma_w^{-1} \vec y 
\label{eq:a_b}
\end{equation}
and
\begin{equation}
    C = \frac{1}{2} \left( W(\vec x) - \vec y \right)^T \Sigma_w^{-1}
    \left( W(\vec x) - \vec y \right)
\label{eq:a_c}
\end{equation}
equation~\eqref{eq:a_loglikelihood}
can be re-written as
\begin{equation}
\ln L(c_i,z,\sigma|\vec y) = -\frac{1}{2} \sum_{i,j} c_i c_j A_{ij} 
+ \sum_i c_i B_i - C
\label{eq:a_simp}
\end{equation}
The most likely set of parameters is that which maximizes the log
likelihood, i.e.,
\begin{equation}
\frac{\partial \ln L(c_i,z,\sigma|\vec y)}{\partial c_k} = -\sum_i c_i A_{ik} + B_k = 0
\label{eq:a_maxlike}
\end{equation}
This equation can readily be solved by inverting the matrix $A_{ik}$, allowing
one to efficiently compute the best-fit linear parameters $c_i$ given a
redshift $z$ and line-width $\sigma$.

Differentiating equation~\eqref{eq:a_maxlike} again yields the Fisher matrix
\begin{equation}
-\frac{\partial^2 \ln L(c_i,z,\sigma|\vec y)}{\partial c_j \partial c_k} = A_{jk} 
\label{eq:a_fisher}
\end{equation}
The parameters of redshift $z$ and line-width $\sigma$ are not included
here: they must be computed separately via non-linear methods. Fortunately,
since the rest of the analysis is linear, the computations for these additional
parameters can be performed in 2-D space. Once we have their best-fitting
values, we can use the inverse of the Fisher matrix to obtain the covariance
matrix $C_{{\rm meas}}$ and to approximately marginalize over any nuisance
parameters. This procedure gives us the line fluxes $\vec c_{{\rm meas}}$ with
a covariance matrix $C_{{\rm meas}}$.

We correct for underlying Balmer continuum 
absorption using the predictions of our best-fitting model to the galaxy's
spectral energy distribution.  This modification of the H$\beta$ flux
may be viewed as a coordinate transformation from the directly measured 
H$\beta$ flux to the corrected flux.  As a result the
Fisher information matrix must be corrected by dividing the row and the
column for H$\beta$ by the correction factor.

Finally, since equation~\eqref{eq:a_likelihood} is quadratic in
the $c_i$ coefficients, the linear parameters $c_i$ have Gaussian errors.

\subsection{Metallicity}
Estimations of galactic oxygen abundances follow from ratios of the line
fluxes. Since metallicity indicators are not defined 
for arbitrary line ratios and since the S/N of emission lines may be of
order 1, simple error propagation from the line fluxes to the line ratios is not
meaningful. Instead, we use multiple metallicity indicators to predict the line
fluxes $\vec{c}_{{\rm pred}}(Z,E,K)$ as a function of metallicity $Z$,
extinction $E=E(B-V)$, and a normalization $K$ to construct a likelihood
function $L(Z,E,K)$ given measured line fluxes. The predicted line fluxes are
\begin{equation}
    \vec c_{{\rm pred}}(Z,E,K)
    = \left(
        \begin{array}{c}
            c_{3727} \\
            c_{3869} \\
            c_{4340} \\
            c_{4861} \\
            c_{5007}
        \end{array}
    \right)
    = \left(
        \begin{array}{c}
            K \, e_{3727}(E) \\
            R_{{\rm Ne3O2}}(Z) \, K \, e_{3869}(E) \\
            0.47 \, c_{4861} \, e_{4340}(E) / e_{4861}(E) \\
            R_{23}^{-1}(Z) \, \left[1 + (1+1/2.98) R_{{\rm O32}}(Z)\right] K \,
                e_{4861}(E) \\
            R_{{\rm O32}}(Z) \, K \, e_{5007}(E)
        \end{array}
    \right)
\label{eq:a_vector}
\end{equation}
where the right hand side is obtained by applying the definitions of the strong
line indicators. Specifically, we allow $K$ to represent the 
[O~II]~$\lambda 3727$ line flux, and use 
the notation $R_i(Z)=10^{r_i(Z)}$ to represent the various polynomial
relations $r_i(Z)$ given in Table~\ref{tab:relations}.  The term
$e_{\lambda}(E)$ gives the extinction and obscuration derived from the
stellar reddening and the \citet{calzetti01} attenuation law.

In our calibration of the strong-line indicators, we account for the scatter
about those relations by measuring the dispersion in the distribution defined
by our local galaxies described in \S\ref{sec:matches} and
\S\ref{sec:calibration}.  As a result, the logarithms of the strong-line ratios
$\log R_i$ have a normal probability distribution at each metallicity $Z$, or
$\log R_i \sim N(r_i(Z),\sigma_{r_i}^2)$. This scatter manifests itself as an
uncertainty in the predicted line fluxes $\vec c_{\rm pred}$. Thus, in addition
to the covariance matrix $C_{{\rm meas}}$ for the measured fluxes, we include a
covariance matrix $C_{{\rm pred}}(Z,E,K) =
\left<\delta{c_i}\,\delta{c_j}\right>$ for the predicted line fluxes. For
example, with the notation of the previous paragraph,
\begin{align}
    \delta c_{3727} &= 0 \\
    \delta c_{3869} &= \delta R_{{\rm Ne3O2}}(Z) \, K \, e_{3869}(E)
    = R_{{\rm Ne3O2}}(Z) \ln{10} \, \delta r_{{\rm Ne3O2}}(Z) \,
    K \, e_{3869}(E) \\
    \left< \delta c_{3727} \delta c_{3727} \right> &= 0 \\
    \left< \delta c_{3727} \delta c_{3869} \right> &= 0 \\
    \left< \delta c_{3869} \delta c_{3869} \right>
    &= \left< \delta r_{\rm Ne3O2}^2(Z) \right> R^2_{\rm Ne3O2}(Z) \ln^2{10} \,
    K^2 \, e^2_{3869}(E) \\
    &= \sigma_{\rm Ne3O2}^2 R^2_{\rm Ne3O2}(Z) \ln^2{10} \, K^2 \, e^2_{3869}(E)
\end{align}
where $\sigma_{\rm Ne3O2}$ is the scatter in the Ne3O2~=~[Ne~III]/[O~II] relation in
Table~\ref{tab:relations}. The other 22 entries in $C_{\rm pred}(Z,E,K) =
\left< \delta c_i \delta c_j \right>$ are obtained in the same fashion.

Finally, we construct the likelihood function. First, given true line fluxes
$\vec c_{\rm true}$, we have a probability distribution for obtaining measured
fluxes $\vec c_{\rm meas}$, and a probability distribution for obtaining
predicted line fluxes $\vec c_{\rm pred}$. We do not know the true line fluxes
$\vec c_{\rm true}$, so we integrate over them to get the likelihood function
\begin{align}
    L(Z,E,K) &= \frac{1}{\sqrt{(2\pi)^5 \det(C_{{\rm meas}} + C_{{\rm pred}})}} \,
    \nonumber\\ &\quad\times
    \exp\left( -\frac{1}{2}
        (\vec c_{{\rm pred}} - \vec{c}_{{\rm meas}})^T
        (C_{{\rm meas}} + C_{{\rm pred}})^{-1}
        (\vec c_{{\rm pred}} - \vec{c}_{{\rm meas}})
    \right)
\label{eq:a_zlikelihood}
\end{align}
We marginalize over $K$ by fitting for the best-fit $K$ at a given $(Z,E)$.
Here, non-linear methods must be used to find the maximum of this likelihood,
as the 1-$\sigma$ confidence intervals are usually too distant from the
best-fit solution for a linear approximation to be valid.

\clearpage

% tab:data
\begin{deluxetable}{cccccccccc}
\tablewidth{0pt}
\tabletypesize{\scriptsize}
\tablecaption{3D-HST Line Fluxes\label{tab:data}}
\tablehead{
\colhead{\textnumero}
&\colhead{$\alpha(2000)$}
&\colhead{$\delta(2000)$}
&\colhead{F140W\tablenotemark{a}}
&\colhead{$z$}
&\colhead{[O~II] $\lambda 3727$\tablenotemark{b,c}}
&\colhead{[Ne~III] $\lambda 3869$\tablenotemark{b,c}}
&\colhead{H$\gamma$\tablenotemark{b,c}}
&\colhead{H$\beta$\tablenotemark{b,c,d}}
&\colhead{[O~III] $\lambda 5007$\tablenotemark{b,c}} }
\startdata
1 & 10:00:18.80 & $+$02:23:08.1 & $23.65$ & $2.09$ & \phantom{0}\ldots & \ldots & \ldots & $2.42 \pm 1.67$ & $\phantom{0}9.88 \pm 1.47$ \\
2 & 10:00:34.79 & $+$02:28:21.5 & $24.54$ & $2.25$ & $\phantom{0}3.09 \pm 1.20$ & $1.14 \pm 1.01$ & $1.64 \pm 0.77$ & \ldots & $\phantom{0}5.94 \pm 0.84$ \\
3 & 10:00:15.40 & $+$02:22:55.0 & $21.89$ & $2.19$ & $11.23 \pm 1.43$ & $2.34 \pm 1.23$ & $1.55 \pm 1.01$ & $8.90 \pm 0.94$ & $22.58 \pm 0.99$ \\
4 & 10:00:23.24 & $+$02:15:55.2 & $23.29$ & $2.09$ & \phantom{0}\ldots & $3.19 \pm 1.50$ & $2.58 \pm 0.97$ & $3.35 \pm 1.24$ & $10.26 \pm 1.58$ \\
5 & 10:00:29.10 & $+$02:17:03.6 & $22.89$ & $1.92$ & $\phantom{0}5.90 \pm 3.76$ & $2.08 \pm 1.76$ & \ldots & $4.19 \pm 0.98$ & $25.93 \pm 0.93$ \\
6 & 10:00:26.72 & $+$02:17:31.2 & $24.46$ & $2.28$ & \phantom{0}\ldots & \ldots & \ldots & $1.94 \pm 0.76$ & $\phantom{0}8.04 \pm 0.90$ \\
7 & 10:00:22.39 & $+$02:17:11.5 & $23.92$ & $2.22$ & \phantom{0}\ldots & \ldots & \ldots & $1.41 \pm 0.82$ & $10.98 \pm 0.97$ \\
8 & 10:00:26.12 & $+$02:17:13.8 & $24.77$ & $2.22$ & \phantom{0}\ldots & \ldots & \ldots & \ldots & $\phantom{0}3.98 \pm 1.07$ \\
9 & 10:00:28.97 & $+$02:13:52.2 & $23.58$ & $2.24$ & $\phantom{0}6.14 \pm 1.34$ & \ldots & \ldots & \ldots & $\phantom{0}7.24 \pm 0.94$ \\
10 & 10:00:24.01 & $+$02:14:09.8 & $23.04$ & $2.10$ & \phantom{0}\ldots & \ldots & $1.23 \pm 1.14$ & $1.54 \pm 1.20$ & $\phantom{0}7.77 \pm 1.17$ \\
\enddata
\tablenotetext{a}{F140W magnitudes.}
\tablenotetext{b}{Flux densities are given in $10^{-17}~\text{ergs}\,\text{s}^{-1}\,\text{cm}^{-2}\,\text{\AA}^{-1}$.}
\tablenotetext{c}{Ellipses denote a signal-to-noise ratio less than 1.0.}
\tablenotetext{d}{H$\beta$ includes the equivalent width correction from SED
fitting.}
\tablecomments{Table 1 is published in its entirety in the electronic
edition of the {\it Astrophysical Journal}.  A portion is shown here
for guidance regarding its form and content.}
\end{deluxetable}

% tab:properties
\begin{deluxetable}{ccccccc}
\tablewidth{0pt}
\tabletypesize{\scriptsize}
\tablecaption{3D-HST Derived Properties\label{tab:properties}}
\tablehead{
\colhead{\textnumero}
&\colhead{Stellar Mass\tablenotemark{a}}
&\colhead{SFR\tablenotemark{b}}
&\colhead{UV slope $\beta$}
&\colhead{$E(B-V)$\tablenotemark{c}}
&\colhead{Metallicity\tablenotemark{d}}
&\colhead{Total Error\tablenotemark{e}}}
\startdata
1 & $\phantom{-}\phantom{0}9.049\pm{}0.048$ & $\phantom{-}0.70\pm{}0.06$ & $-2.25\pm{}0.06$ & \ldots & $8.26_{<7.0}^{8.97}$ & \ldots \\
2 & $\phantom{-}\phantom{0}8.927\pm{}0.096$ & $\phantom{-}0.46\pm{}0.12$ & $-2.12\pm{}0.13$ & $0.063\pm{}0.061$ & $8.28_{8.03}^{8.65}$ & $0.62$ \\
3 & $\phantom{-}10.142\pm{}0.019$ & $\phantom{-}2.18\pm{}0.02$ & $-1.00\pm{}0.02$ & $0.601\pm{}0.011$ & $8.58_{8.33}^{8.80}$ & $0.47$ \\
4 & $\phantom{-}\phantom{0}9.099\pm{}0.061$ & $\phantom{-}1.29\pm{}0.05$ & $-1.57\pm{}0.05$ & $0.327\pm{}0.025$ & \ldots & \ldots \\
5 & $\phantom{-}\phantom{0}9.070\pm{}0.035$ & $\phantom{-}1.13\pm{}0.03$ & $-1.86\pm{}0.03$ & $0.188\pm{}0.015$ & $7.96_{7.59}^{8.19}$ & $0.60$ \\
6 & $\phantom{-}\phantom{0}8.375\pm{}0.113$ & $\phantom{-}0.71\pm{}0.06$ & $-2.08\pm{}0.07$ & $0.080\pm{}0.033$ & \ldots & \ldots \\
7 & $\phantom{-}\phantom{0}8.743\pm{}0.100$ & $\phantom{-}0.94\pm{}0.09$ & $-1.71\pm{}0.10$ & $0.258\pm{}0.047$ & $7.78_{<7.0}^{8.14}$ & \ldots \\
8 & $\phantom{-}\phantom{0}7.735\pm{}0.082$ & $\phantom{-}0.04\pm{}0.11$ & $-2.65\pm{}0.11$ & \ldots & $7.24_{<7.0}^{7.83}$ & \ldots \\
9 & $\phantom{-}\phantom{0}9.044\pm{}0.034$ & $\phantom{-}1.01\pm{}0.04$ & $-1.96\pm{}0.05$ & $0.139\pm{}0.022$ & $8.61_{8.38}^{8.88}$ & $0.50$ \\
10 & $\phantom{-}\phantom{0}9.462\pm{}0.024$ & $\phantom{-}1.52\pm{}0.05$ & $-1.32\pm{}0.05$ & $0.448\pm{}0.026$ & $8.35_{<7.0}^{8.93}$ & \ldots \\
\enddata
\tablenotetext{a}{Units are $\log{M/M_{\odot}}$.}
\tablenotetext{b}{Units are $\log{{\rm SFR}/M_{\odot}}$~yr$^{-1}$.}
\tablenotetext{c}{Ellipses denote where the UV slope gives a negative
obscuration.  For these objects, we set $E(B-V)=0$.}
\tablenotetext{d}{Metallicity means $12+\log({\rm O/H})$; upper and 
lower limits are given as suffixes.}
\tablenotetext{e}{Total error in metallicity.}
\tablecomments{Table 2 is published in its entirety in the electronic 
edition of the {\it Astrophysical Journal}.  A portion is shown here 
for guidance regarding its form and content.}
\end{deluxetable}

% tab:relations
\begin{deluxetable}{crrrc}
\tablewidth{0pt}
\tabletypesize{\scriptsize}
\tablecaption{High-EW Strong-Line Ratio Relations Coefficients\label{tab:relations}}
\tablehead{
\colhead{Relation $R_i$}
&\colhead{1}
&\colhead{$x$}
&\colhead{$x^2$}
&\colhead{Scatter $\sigma_{r_i}$}}
\startdata
$R_{23}$            & $ 0.7675$ & $-0.4861$ & $-0.3169$ & $0.03771$ \\
{[O~III]/H$\beta$}  & $ 0.3929$ & $-0.7610$ & $-0.3669$ & $0.06760$ \\
{[O~II]/H$\beta$}   & $ 0.3922$ & $-0.2275$ & $-0.8205$ & $0.10521$ \\
{[N~II]/H$\alpha$}  & $-0.7798$ & $ 1.0752$ &           & $0.13512$ \\
{[O~III]/[O~II]}    & $-0.1149$ & $-1.0173$ &           & $0.16025$ \\
{[O~III]/[N~II]}    & $ 0.8107$ & $-1.4442$ &           & $0.17676$ \\
{[Ne~III]/[O~II]}   & $-1.2118$ & $-0.9809$ &           & $0.14452$
\enddata
\tablecomments{We use the \citet{maiolino+08} convention $x=12+\log({\rm O/H})
- 8.69$.  Each relation is modelled using $\log R_i = r_i(x) +
w(\sigma_{r_i})$, where $r_i(x)$ is a polynomial with the coefficients
indicated in the table, and $w(\sigma_{r_i})$ is Gaussian white noise with
scatter $\sigma_{r_i}$. Put another way, $\log R_i$ is modeled as a random
variable distributed as $\log{R_i} \sim N(r_i(x),\sigma_{r_i}^2)$. These
have been calibrated using our local sample of galaxies matched in stellar mass
and SFR (see \S\ref{sec:matches} and \S\ref{sec:calibration}), and we plot them
in Figure~\ref{fig:newmaiolino}. Here, [O~III] refers to the forbidden line at
5007~\AA, [O~II] to $\lambda 3727$, [N~II] to $\lambda 6584$, Ne~[III] $\lambda
3869$, and $R_{23}$ to ([O~II] $\lambda 3727$ + [O~III] $\lambda\lambda
4959,5007$)/H$\beta$.}
\end{deluxetable}

\end{document}

%% file: zcompare_matched.tex
\begin{tikzpicture}
    \begin{axis}[
            scale=1.3,
            axis equal image,
            xlabel={$12+\log({\rm O/H})$ [``$T_e$'' method]},
            ylabel={$12+\log({\rm O/H})$ [Equations from Table~\ref{tab:relations}]},
            clip marker paths=true,
        ]
        \addplot[only marks, fill=gray, mark size=0.5pt]
        table[x=Z_Te_OIII,y=Z_weighted_matched]{data_matches_sn4363.dat};
        \addplot[dashed, ultra thick, domain=7.0:9.0, red]{x};
    \end{axis}
\end{tikzpicture}

%% file: ms.bbl
\clearpage

\begin{thebibliography}{}

\bibitem[Acquaviva \etal(2011)]{acq+11} Acquaviva, V., 
Gawiser, E., \& Guaita, L. 2011, \apj, 737, 47

\bibitem[Acquaviva \etal(2012)]{acq+12} Acquaviva, V., private communication

\bibitem[Andrews \& Martini(2013)]{andrews+13} Andrews, B.H., \& Martini, P. 
2013, \apj, 765, 140 

\bibitem[Asplund \etal(2009)]{asplund+09} Asplund, M., Grevesse, N., 
Sauval, A.J., \& Scott, P. 2009, \araa, 47, 481 

\bibitem[Atek \etal(2011)]{atek+11} Atek, H., Siana, B., 
Scarlata, C., \etal\  2011, \apj, 743, 121 

\bibitem[Baldwin \etal(1981)]{bpt81} Baldwin, J.A., Phillips, M.M., \& 
Terlevich, R. 1981, \pasp, 93, 5 

\bibitem[Belli \etal(2013)]{belli+13} Belli, S., Jones, T., 
Ellis, R.S., \& Richard, J. 2013, \apj, 772, 141 

\bibitem[Brammer \etal(2012)]{brammer+12} Brammer, G.B., van Dokkum, P.G., 
Franx, M., \etal\  2012, \apjs, 200, 13 

\bibitem[Brinchmann et al.(2004)]{brinchmann+04} Brinchmann, J., 
Charlot, S., White, S.D.M., \etal\  2004, \mnras, 351, 1151

\bibitem[Brinchmann \etal(2008)]{brinchmann+08} Brinchmann, J., Pettini, M., 
\& Charlot, S. 2008, \mnras, 385, 769

\bibitem[Brooks \& Gelman(1998)]{brooks+98} Brooks, S., \& Gelman, A. 1998, 
J.~Comp.~\& Graph.~Stat., 7, 434

\bibitem[Bruzual \& Charlot(2003)]{BC03} Bruzual, G., \& Charlot, S. 
2003, \mnras, 344, 1000 

\bibitem[Buat \etal(2012)]{buat+12} Buat, V., Noll, S., Burgarella, D., 
\etal\  2012, \aap, 545, A141

\bibitem[Cardelli \etal(1989)]{cardelli+89} Cardelli, J.A., 
Clayton, G.C., \& Mathis, J.S. 1989, \apj, 345, 245 

\bibitem[Calzetti(2001)]{calzetti01} Calzetti, D. 2001, \pasp, 113, 1449 

\bibitem[Chabrier(2003)]{chabrier03} Chabrier, G. 2003, \pasp, 115, 763 

\bibitem[Charlot \& Fall(2000)]{charlot+00} Charlot, S., \& Fall, S.M.
2000, \apj, 539, 718 

\bibitem[Coil \etal(2015)]{coil+15} Coil, A.L., Aird, J., 
Reddy, N., \etal\ 2015, \apj, 801, 35

\bibitem[Conroy(2013)]{conroy13} Conroy, C. 2013, \araa, 51, 393 

\bibitem[Cullen \etal(2014)]{cullen+14} Cullen, F., Cirasuolo, M.,
McLure, R.J., Dunlop, J.S., \& Bowler, R.A.A. 2014, \mnras, 440, 2300 

\bibitem[Dav{\'e} \etal(2012)]{dave+12} Dav{\'e}, R., 
Finlator, K., \& Oppenheimer, B.D.  2012, \mnras, 421, 98 

\bibitem[Dayal \etal(2009)]{dayal+09} Dayal, P., Ferrara, A., 
Saro, A., \etal\  2009, \mnras, 400, 2000

\bibitem[Dayal \etal(2013)]{dayal+13} Dayal, P., Ferrara, A., 
\& Dunlop, J.S. 2013, \mnras, 430, 2891 

\bibitem[de los Reyes \etal(2015)]{delosreyes+15} de los Reyes, 
M.A., Ly, C., Lee, J.C., \etal\  2015, \aj, 149, 79

\bibitem[Dickinson \etal(2003)]{dickinson+03} Dickinson, M., 
Papovich, C., Ferguson, H.C., \& Budav\'ari, T. 2003, \apj, 587, 25 

\bibitem[Driver \etal(2011)]{driver+11} Driver, S.P., Hill, D.T.,
Kelvin, L.S., \etal\  2011, \mnras, 413, 971

\bibitem[Erb \etal(2006)]{erb+06a} Erb, D.K., Shapley, A.E., 
Pettini, M., \etal\ 2006a, \apj, 644, 813 

\bibitem[Erb \etal (2006)]{erb+06b} Erb, D.K., Steidel, C.C., 
Shapley, A.E., \etal\ 2006b, \apj, 647, 128 

\bibitem[Finkelstein \etal(2011)]{finkelstein+11} Finkelstein, S.L., 
Cohen, S.H., Moustakas, J., \etal\ 2011, \apj, 733, 117 

\bibitem[Forbes \etal(2014)]{forbes+14} Forbes, J.C., Krumholz, M.R.,
Burkert, A., \& Dekel, A. 2014, \mnras, 443, 168 

\bibitem[F{\"o}rster Schreiber \etal(2009)]{forster+09} 
F{\"o}rster Schreiber, N.M., Genzel, R., Bouch\'e, N., \etal\ 2009, 
\apj, 706, 1364 

\bibitem[Fruchter \etal(2009)]{fruchter+09} Fruchter, A., Sosey, M.,
Hack, W., \etal\ 2009, The MultiDrizzle Handbook Version 3.0 
(Baltimore: STScI)

\bibitem[Garn \& Best(2010)]{garn-best10} Garn, T., \& Best, P.N. 2010, 
\mnras, 409, 421 

\bibitem[Gelman \& Rubin(1992)]{gelman+92} Gelman, A., \& Rubin, D. 1992, 
Stat.~Sci., 7, 457

\bibitem[Giavalisco \etal(2004)]{GOODS} Giavalisco, M., Ferguson, H.C.,
Koekemoer, A.M., \etal\  2004, \apjl, 600, L93

\bibitem[Grogin \etal(2011)]{CANDELS} Grogin, N.A., Kocevski, D.D.,
Faber, S.M., \etal\ 2011, \apjs, 197, 35 

\bibitem[Hagen \etal(2014)]{hagen+14} Hagen, A., Ciardullo, R., 
Gronwall, C., \etal\  2014, \apj, 786, 59 

\bibitem[Hao \etal(2011)]{hao+11} Hao, C.-N., Kennicutt, R.C.,
Johnson, B.D., \etal\  2011, \apj, 741, 124 

\bibitem[Henry \etal(2013)]{henry+13} Henry, A., Scarlata, C., 
Dom{\'{\i}}nguez, A., \etal\  2013, \apjl, 776, L27 

\bibitem[Hinshaw \etal(2013)]{hinshaw+13} Hinshaw, G., Larson, D.,
Komatsu, E., \etal\  2013, \apjs, 208, 19

\bibitem[Holden \etal(2014)]{holden+14} Holden, B.P., Oesch, P.A., 
Gonzalez, V.G., \etal\  2014, submitted to ApJ (arXiv:1401.5490)

\bibitem[Hopkins \& Beacom(2006)]{hopkins+06} Hopkins, A.M., \& Beacom, J.F.
2006, \apj, 651, 142 

\bibitem[Horne(1986)]{horne86} Horne, K. 1986, \pasp, 98, 609

\bibitem[Hummer \& Storey(1987)]{hummer+87} Hummer, D.G., \& Storey, P.J.
1987, \mnras, 224, 801

\bibitem[Izotov \etal(2006)]{izotov+06} Izotov, Y.I., Stasi{\'n}ska, G., 
Meynet, G., Guseva, N.G., \& Thuan, T.X. 2006, \aap, 448, 955 

\bibitem[Izotov \etal(2011)]{izotov+11} Izotov, Y.I., Guseva, N.G., \& 
Thuan, T.X. 2011, \apj, 728, 161

\bibitem[Juneau \etal(2014)]{juneau+14} Juneau, S., Bournaud, F., Charlot, S., 
\etal\ 2014, \apj, 788, 88 

\bibitem[Kashino \etal(2013)]{kashino+13} Kashino, D., Silverman, J.D.,
Rodighiero, G., \etal\  2013, \apjl, 777, L8

\bibitem[Kauffmann et al.(2003)]{kauffmann+03} Kauffmann, G., 
Heckman, T.~M., Tremonti, C., et al.\ 2003, \mnras, 346, 1055

\bibitem[Kennicutt \& Evans(2012)]{kennicutt+12} Kennicutt, R.C., \& 
Evans, N.J. 2012, \araa, 50, 531 

\bibitem[Kewley \& Dopita(2002)]{kewley+02} Kewley, L.J., \& Dopita, M.A.
2002, \apjs, 142, 35 

\bibitem[Kewley \& Ellison(2008)]{kewley+08} Kewley, L.J., \& Ellison, S.L.
2008, \apj, 681, 1183 

\bibitem[Kewley \etal(2013)]{kewley+13} Kewley, L.J., Maier, C.,
Yabe, K., \etal\  2013, \apjl, 774, L10 

\bibitem[Kimble \etal(2008)]{kimble+08} Kimble, R.A., MacKenty, J.W.,
O'Connell, R.W., \& Townsend, J.A. 2008, \procspie, 7010, 70101E

\bibitem[Koekemoer \etal(2011)]{koekemoer+11} Koekemoer, A.M., 
Faber, S.M., Ferguson, H.C., \etal\  2011, \apjs, 197, 36

\bibitem[Kroupa(2001)]{kroupa01} Kroupa, P. 2001, \mnras, 322, 231

\bibitem[Kulas \etal(2013)]{kulas+13} Kulas, K.R., McLean, I.S.,
Shapley, A.E., \etal\  2013, \apj, 774, 130 

\bibitem[K{\"u}mmel \etal(2009)]{kummel+09} K{\"u}mmel, M., Walsh, J.R., 
Pirzkal, N., Kuntschner, H., \& Pasquali, A. 2009, \pasp, 121, 59 

\bibitem[Lara-L\'opez \etal(2010)]{lara-lopez+10} Lara-L\'opez, M.A., 
Cepa, J., Bongiovanni, A., \etal\ 2010, \aap, 521, L53 

\bibitem[Lara-L\'opez \etal(2013)]{lara-lopez+13} Lara-L\'opez, M.A.,
Hopkins, A.M., L\'opez-S\'anchez, A.R., \etal\  2013, \mnras, 434, 451 

\bibitem[Lequeux \etal(1979)]{lequeux+79} Lequeux, J., Peimbert, M., 
Rayo, J.F., Serrano, A., \& Torres-Peimbert, S. 1979, \aap, 80, 155 

\bibitem[Levesque \& Richardson(2014)]{levesque+14} Levesque, E.M., \& 
Richardson, M.L.A. 2014, \apj, 780, 100 

\bibitem[Lewis \& Bridle(2002)]{lewis+02} Lewis, A., \& Bridle, S. 2002, 
\prd, 66, 103511 

\bibitem[Lilly \etal(2013)]{lilly+13} Lilly, S.J., Carollo, C.M.,
Pipino, A., Renzini, A., \& Peng, Y. 2013, \apj, 772, 119 

\bibitem[Madau(1995)]{madau95} Madau, P. 1995, \apj, 441, 18 

\bibitem[Madau \& Dickinson(2014)]{madau+14} Madau, P., \& Dickinson, M. 2014, 
\araa, 52, 415

\bibitem[Maier \etal(2014)]{maier+14} Maier, C., Lilly, S.J., 
Ziegler, B., \etal\ 2014, \apj, 792, 3

\bibitem[Maiolino \etal(2008)]{maiolino+08} Maiolino, R., Nagao, T., 
Grazian, A., \etal\  2008, \aap, 488, 463 

\bibitem[Mannucci \etal(2009)]{mannucci+09} Mannucci, F., Cresci, G., 
Maiolino, R., \etal\ 2009, \mnras, 398, 1915 

\bibitem[Mannucci \etal(2010)]{mannucci+10} Mannucci, F., Cresci, 
G., Maiolino, R., Marconi, A., \& Gnerucci, A. 2010, \mnras, 408, 2115 

\bibitem[Mannucci \etal(2011)]{mannucci+11} Mannucci, F., 
Salvaterra, R., \& Campisi, M.A. 2011, \mnras, 414, 1263 

\bibitem[Murphy \etal(2011)]{murphy+11} Murphy, E.J., Condon, J.J.,
Schinnerer, E., \etal\  2011, \apj, 737, 67 

\bibitem[Nagao \etal(2006)]{nagao+06} Nagao, T., Maiolino, R., \& 
Marconi, A. 2006, \aap, 459, 85

\bibitem[Nakajima \etal(2013)]{nakajima+13} Nakajima, K., Ouchi, M.,
Shimasaku, K., \etal\  2013, \apj, 769, 3 

\bibitem[Nakajima \& Ouchi(2014)]{nakajima+14} Nakajima, K., \& Ouchi, M. 
2014, \mnras, 442, 900 

\bibitem[Oey \& Shields(2000)]{oey+00} Oey, M.S., \& Shields, J.C. 2000, 
\apj, 539, 687 

\bibitem[Osterbrock \& Ferland(2006)]{AGN3} Osterbrock, D.E., \& Ferland, G.J. 
2006, Astrophysics of Gaseous Nebulae and Active Galactic Nuclei, 2nd.~ed.~by 
D.E. Osterbrock \& G.J. Ferland (Sausalito, CA: University Science Books)

\bibitem[Pacifici \etal(2013)]{pacifici+13} Pacifici, C., Kassin, S.A.,
Weiner, B., Charlot, S., \& Gardner, J.P. 2013, \apjl, 762, L15

\bibitem[Pagel \etal(1979)]{pagel+79} Pagel, B.E.J., Edmunds, M.G.,
Blackwell, D.E., Chun, M.S., \& Smith, G. 1979, \mnras, 189, 95 

\bibitem[Pettini \& Pagel(2004)]{pettini+04} Pettini, M., \& 
Pagel, B.E.J.  2004, \mnras, 348, L59 

\bibitem[Pipino \etal(2014)]{pipino+14} Pipino, A., Lilly, S.J.,
\& Carollo, C.M.  2014, \mnras, 441, 1444 

\bibitem[Planck Collaboration \etal(2014)]{planck+14} Planck
Collaboration, Ade, P.A.R., Aghanim, N., \etal\  2014, \aap, 571, AA16

\bibitem[Price \etal(2014)]{price+14} Price, S.H., Kriek, M., 
Brammer, G.B., \etal\ 2014, \apj, 788, 86 

\bibitem[Reddy \etal(2012)]{reddy+12} Reddy, N.A., Pettini, M.,
Steidel, C.C., \etal\  2012, \apj, 754, 25

\bibitem[Reddy \etal(2015)]{reddy+15} Reddy, N.A., Kriek, M., 
Shapley, A.E., \etal\  2015, \apj, 806, 259

\bibitem[Salmon \etal(2015)]{salmon+15} Salmon, B., Papovich, C.,
Finkelstein, S.L., \etal\  2015, \apj, 799, 183

\bibitem[Salpeter(1955)]{salpeter55} Salpeter, E.E. 1955, \apj, 121, 161 

\bibitem[Sanders \etal(2015)]{sanders+15} Sanders, R.L., 
Shapley, A.E., Kriek, M., \etal\  2015, \apj, 799, 138

\bibitem[Schaerer \& de Barros(2009)]{schaerer+09} Schaerer, D., \& 
de Barros, S. 2009, \aap, 502, 423 

\bibitem[Scoville \etal(2007)]{COSMOS} Scoville, N., Aussel, H.,
Brusa, M., \etal\  2007, \apjs, 172, 1

\bibitem[Scoville \etal(2015)]{scoville+15} Scoville, N., Faisst, A.,
Capak, P., \etal\  2015, \apj, 800, 108

\bibitem[Shapley \etal(2015)]{shapley+15} Shapley, A.E., Reddy, N.A.,
Kriek, M., \etal\ 2015,  \apj, 801, 88

\bibitem[Shirazi \etal(2014)]{shirazi+14} Shirazi, M., Brinchmann, J., \& 
Rahmati, A. 2014, \apj, 787, 120 

\bibitem[Skelton \etal(2014)]{skelton+14} Skelton, R.E., 
Whitaker, K.E., Momcheva, I.G., \etal\  2014, \apjs, 214, 24

\bibitem[Song \etal(2014)]{song+14} Song, M., Finkelstein, S.L.,
Gebhardt, K., \etal\  2014, \apj, 791, 3 

\bibitem[Steidel \etal(2014)]{steidel+14} Steidel, C.C., Rudie, G.C.,
Strom, A.L., \etal\  2014, \apj, 795, 165

\bibitem[Storey \& Zeippen(2000)]{storey+00} Storey, P.J., \& Zeippen, C.J.
2000, \mnras, 312, 813 

\bibitem[Tielens(2008)]{tielens08} Tielens, A.G.G.M. 2008, \araa, 46, 289 

\bibitem[Tremonti \etal(2004)]{tremonti+04} Tremonti, C.A., 
Heckman, T.M., Kauffmann, G., \etal\ 2004, \apj, 613, 898 

\bibitem[Trump \etal(2013)]{trump+13} Trump, J.R., Konidaris, N.P.,
Barro, G., \etal\  2013, \apjl, 763, L6 

\bibitem[Whitaker \etal(2014)]{whitaker+14} Whitaker, K.E., Franx, M., 
Leja, J., \etal\  2014, \apj, 795, 104 

\bibitem[Weiner \& the AGHAST Team(2014)]{weiner+14} Weiner, B.J., \& AGHAST 
Team 2014, \baas, 223, \#227.07

\bibitem[Wuyts \etal(2013)]{wuyts+13} Wuyts, S., F\"orster Schreiber, N.M., 
Nelson, E.J., \etal\ 2013, \apj, 779, 135 

\bibitem[Wuyts \etal(2014)]{wuyts+14} Wuyts, E., Kurk, J., 
F{\"o}rster Schreiber, N.M., \etal\  2014, \apjl, 789, L40 

\bibitem[Yates \etal(2012)]{yates+12} Yates, R.M., Kauffmann, G.,
\& Guo, Q. 2012, \mnras, 422, 215 

\bibitem[Zahid \etal(2014)]{zahid+14} Zahid, H.J., Kashino, D.,
Silverman, J.D., \etal\  2014, \apj, 792, 75 

\bibitem[Zeimann \etal(2014)]{zeimann+14} Zeimann, G.R., 
Ciardullo, R., Gebhardt, H., \etal\  2014, \apj, 790, 113

\end{thebibliography}
